\documentclass[letterpaper]{article} 
\usepackage[submission]{aaai25}  
\usepackage{times}  
\usepackage{comment}
\usepackage{helvet}  
\usepackage{courier}  
\usepackage[hyphens]{url}  
\usepackage{graphicx} 
\usepackage{soul}
\usepackage{natbib}
\usepackage{natbib}  
\usepackage{caption} 
\usepackage{algorithm}
\usepackage{algorithmic}
\usepackage{booktabs}
\usepackage{subfigure}

\usepackage{newfloat}
\usepackage{listings}
\usepackage{multirow}

\usepackage{xcolor}

\newcommand{\answerYes}[1]{\textcolor{blue}{#1}} 
\newcommand{\answerNo}[1]{\textcolor{teal}{#1}} 
\newcommand{\answerNA}[1]{\textcolor{gray}{#1}}

\DeclareCaptionStyle{ruled}{labelfont=normalfont,labelsep=colon,strut=off} 
\floatstyle{ruled}
\newfloat{listing}{tb}{lst}{}
\floatname{listing}{Listing}
\makeatletter
\gdef\showauthors@on{T}
\gdef\copyright@on{T}%
\long\gdef\pdfinfo #1{\relax}%
\makeatother

\title{Large Engagement Networks for Classifying Coordinated Campaigns and Organic Twitter Trends}

\author {
    Atul Anand Gopalakrishnan\textsuperscript{\rm 1},
    Jakir Hossain\textsuperscript{\rm 1},
    Tu\u{g}rulcan Elmas\textsuperscript{\rm 2},
    Ahmet Erdem Sar{\i}y\"{u}ce\textsuperscript{\rm 1}
}
\affiliations {
    \textsuperscript{\rm 1}University at Buffalo\\
    \textsuperscript{\rm 2} University of Edinburgh\\
    atulanan@buffalo.edu, mh267@buffalo.edu,
    telmas@ed.ac.uk, erdem@buffalo.edu
}

\usepackage{bibentry}

\begin{document}

\maketitle

\begin{abstract}



Social media users and inauthentic accounts, such as bots, may coordinate in promoting their topics. Such topics may give the impression that they are organically popular among the public, even though they are astroturfing campaigns that are centrally managed. It is challenging to predict if a topic is organic or a coordinated campaign due to the lack of reliable ground truth. In this paper, we create such ground truth by detecting the campaigns promoted by ephemeral astroturfing attacks. These attacks push any topic to Twitter's (X) trends list by employing bots that tweet in a coordinated manner in a short period and then immediately delete their tweets. We manually curate a dataset of organic Twitter trends. We then create engagement networks out of these datasets which can serve as a challenging testbed for graph classification task to distinguish between campaigns and organic trends. Engagement networks consist of users as nodes and engagements as edges (retweets, replies, and quotes) between users. We release the engagement networks for 179 campaigns and 135 non-campaigns, and also provide finer-grain labels to characterize the type of the campaigns and non-campaigns. Our dataset, LEN (Large Engagement Networks), is available in the URL below. In comparison to traditional graph classification datasets, which are small with tens of nodes and hundreds of edges at most, graphs in LEN are larger. The average graph in LEN has $\sim$11K  nodes and $\sim$23K edges. We show that state-of-the-art GNN methods give only mediocre results for campaign vs. non-campaign and campaign type classification on LEN. LEN offers a unique and challenging playfield for the graph classification problem. We believe that LEN will help advance the frontiers of graph classification techniques on large networks and also provide an interesting use case in terms of distinguishing coordinated campaigns and organic trends.

\end{abstract}
\begin{links}
  \link{Code}{https://github.com/erdemUB/LEN}
  \link{Datasets}{https://erdemub.github.io/large-engagement-network/}
\end{links}

\section{Introduction}
Social media serves as a censor to public sentiment, reflecting popular topics of widespread interest through organic discussions among users. For instance, Twitter (recently renamed as X) monitors popular topics, trends, and publishes them on its main page, implying that those are the topics that users widely discuss. On the other hand, coordinated efforts can manipulate perceptions on certain topics. Users with common goals may attempt to artificially inflate the popularity of certain topics to promote their campaigns. They may employ fake accounts and bots in a coordinated manner to achieve that while hiding those accounts' inauthentic nature, which is a strategy named astroturfing~\citep{elmas2021ephemeral}. Such efforts can obscure genuine discourse, presenting a challenge in discerning topics that are popular due to organic activity from coordinated campaigns. Twitter's trends are also susceptible to such manipulation. Past studies reported that adversaries manipulate Twitter trends frequently in various countries, such as Pakistan~\citep{kausar2021towards}, India~\citep{jakesch2021trend}, and Turkey~\citep{elmas2021ephemeral}.

We focus on the latter case, where the adversaries primarily employ a special attack named ``ephemeral astroturfing''. In this attack, a set of bots promote a topic (a hashtag or an n-gram representing a campaign) by bulk-tweeting it in a text that is randomly generated using a lexicon. They then immediately delete their tweets. Despite this, the topics still appear on the trend lists. Since this attack is both effective and easy to detect due to its distinct activity pattern, it helps us to establish a reliable ground truth on the topics that are campaigns.

Our work aims to create a graph classification benchmark of Turkish Twitter engagement networks to help identify campaign graphs and other downstream tasks (such as identifying the type of campaign). To do this, we detect ephemeral astroturfing attacks and annotate their target topics as campaigns. Manual verification of these annotations shows that they are mostly related to politics, financial promotions (e.g., cryptocurrencies), and groups of people organizing themselves to call for reforms. The collected data is then converted to a set of engagement networks or graphs, where the nodes are the users and the edges indicate engagements between the users, which in our case can be retweets, replies, or quotes. Our dataset, LEN, contains 314 large networks, 179 campaign and 135 non-campaign, containing 11,769 nodes and 23,593 edges, on average. We further provide finer-grain labels for the types of campaigns and non-campaigns.
LEN is publicly available at \url{https://erdemub.github.io/large-engagement-network/}.
The dataset is released under a CC-BY license, enabling free sharing and adaptation for research or development purpose.

In the rest, we first provide a background and summarize related works on graph classification methods, graph classification datasets, and trend manipulation. Then we provide a detailed description of how the data is collected from Twitter and converted into graphs for classification tasks. Next, we conduct graph classification experiments on LEN using established GNNs, performing both campaign vs. non-campaign classification and campaign type detection.
Finally, we discuss the limitations and ethic of our dataset. LEN offers a challenging testbed for the graph classification problem. We believe that our dataset will help advance the frontiers of graph classification techniques on large networks and also provides an interesting use case in terms of distinguishing coordinated campaigns and organic trends.

\section{Related work}
\label{sec:relwork}
In this section, we first provide a brief overview of graph classification methods, and then summarize the datasets tailored for this task. We also discuss recent studies on trend manipulation.

\subsection{Graph classification methods}

Graph classification is a fundamental task in machine learning with applications in bioinformatics, chemistry, social network analysis, and malware detection~\citep{lee2018graph, you2020graph,wu23}.
At high level, an embedding is created for each graph in a given dataset and then those embeddings are used for classification.
There are broadly two approaches for graph classification, namely graph kernels and graph neural networks (GNNs). Graph kernels measure the similarities between each pair of graphs, using similarity functions that compare structural properties. A kernel matrix is constructed using the pairwise similarities between all graphs. This matrix is then fed to a kernel-based machine learning model (e.g., SVMs) for graph classification. Different approaches exist, primarily distinguished by the kernel function employed. 
The methods include random-walk based approaches~\citep{hammack2011handbook, kang2012fast, sugiyama2015halting}, shortest-path based approaches~\citep{borgwardt2005shortest}, graph-matching~\citep{duchenne2011graph, frohlich2005assignment}, neighbourhood based approaches~\citep{shervashidze2011weisfeiler, morris2017glocalized}, and graphlet-based methods~\citep{shervashidze2009efficient}. 

A major drawback of kernel-based approaches is the inability to learn feature extraction and the downstream classification task simultaneously.
GNNs overcome this issue thanks to neural network architectures, which automatically create features by using message-passing~\citep{kipf2016semi}.
Here, each node has an embedding and sends it as a message to all the neighboring nodes. Each node then aggregates the messages from neighbors and updates its embedding. Over the years, there have been many approaches to aggregating neighborhood embeddings. GCN uses dual-degree normalization to account for the varying number of neighbors each node may have~\citep{kipf2016semi}, GAT uses attention-weight to assign varying weights to each neighbour~\citep{velivckovic2018graph}, and GIN uses an MLP to perform aggregation using a trainable parameter ($\epsilon$) to determine the amount of importance given to the ego node in comparison to its neighbours~\citep{xu2018powerful}. To obtain a graph-level embedding, the node embeddings are pooled. The simplest way to do this is via a simple readout function like Max-Pool or Average-Pool. However, due to the structural properties of graphs, a readout function does not preserve structural knowledge about the graph. More effective pooling methods include SORTPOOL, which sorts the nodes using its WL-color obtained from the final layer of applying a GNN~\citep{zhang2018end}, and Hierarchical pooling methods which focus on coarsening the graph after message-passing to capture structural information about the graph~\citep{ying2018hierarchical,bianchi2020hierarchical,bianchi2020spectral,bacciu2023graph,lee2019self}. GNNs tend to falter while capturing global information and long-range dependencies, often leading to issues like over-smoothing and over-squashing~\citep{alon2020bottleneck,topping2021understanding}.  In this paper we use average pooling because our primary motive is to understand how graph ML models performs with respect to our dataset.
\subsection{Graph classification datasets}
Given the importance of graph classification, several datasets have been curated within various application domains. 
Table \ref{tab:dataset_description} shows a summary of established graph classification datasets. 


\begin{table}[t]
    \setlength{\tabcolsep}{1mm}
    \fontsize{9pt}{11pt}\selectfont
    \centering
    \begin{tabular}{l|l|l|l|l|l}
    \multirow{2}{*}{\fontsize{10}{12}\selectfont \textbf{Categ.}}  & \fontsize{10}{12}\selectfont \textbf{Dataset} & \fontsize{10}{12}\selectfont \textbf{\#} & \fontsize{10}{12}\selectfont \textbf{Avg.} & \fontsize{10}{12}\selectfont \textbf{Avg.} & \fontsize{10}{12}\selectfont \textbf{\#}\\ 
    & & \fontsize{10}{12}\selectfont \textbf{graphs} & \fontsize{10}{12}\selectfont \textbf{\# nodes} & \fontsize{10}{12}\selectfont \textbf{\# edges} & \fontsize{10}{12}\selectfont \textbf{classes}\\

    \hline
    \multirow{8}{*}{\rotatebox[origin=c]{90}{Biological}} & MUTAG & 118 & 17.9 & 20 & 2\\
                                        & PTC-FR & 349 & 14.11 & 14.48 & 2\\
                                        & PTC-MR & 344 & 14.29 & 14.69 & 2 \\
                                        & PTC-FM & 349 & 14.11 & 14.48 & 2\\
                                        & PTC-MM & 336 & 13.97 & 14.32 & 2\\
                                        & NCI1 & 4110 & 29.8 & 64.69 & 2\\
                                        & ENZYMES & 600 & 32.63 & 62.14 & 6\\
                                        & PROTEINS & 1113 & 39.06 & 72.82 & 2\\
                                        & obgn-molhiv & 41,127 & 25.5 & 27.5  & 2\\
                                        & obgn-molpcba & 437,929 & 26.0 & 28.1 & 2\\
                                        & obgn-ppa & 158,100 & 243.4 & 2,266.1 & 37\\
                                        \hline
    \multirow{6}{*}{\rotatebox[origin=c]{90}{Social}} & IMDB-B & 1000 & 19.77 & 96.53 & 2\\
                                                        & IMDB-M & 1500 & 13 & 65.94 & 3\\
                                                        & REDDIT-B & 2000 & 429.63 & 497.75 & 2\\
                                                        & REDDIT-M-5K & 4999 & 508.52 & 594.87 & 5\\
                                                        & REDDIT-M-12K & 11929 & 391.41 & 456.89 & 11\\
                                                        & COLLAB & 5000 & 74.49 & 2457.78 & 3\\ \hline
    Misc. & MalNet & 1.2M & 15,378  & 35,167 & 696\\ \hline
    \multirow{2}{*}{\bf Ours} & {\bf Small} & {\bf 100}& {\bf 2,070.63 }& {\bf  2,696.23}& {\bf  13}\\
    & {\bf Original} & {\bf 314} & {\bf 11,769.23}&  {\bf 23,593.97 }& {\bf 15 }\\
    \end{tabular}
    \caption{Comparison of graph classification datasets to our large engagement networks.}
    \label{tab:dataset_description}
\vspace{-2ex}
\end{table}

Biological datasets are typically either molecule-based graphs and protein graphs. Molecule graphs (MUTAG, PTC, and NCI1) are labeled based on bioinformatics applications such as disease-curing effectiveness~\citep{kriege2012subgraph, shervashidze2011weisfeiler}. 
MUTAG consists of compound graphs with binary labels that indicate if they are effective against the Salmonella. PTC are molecule graphs extracted from rodents, labeled with one of eight levels of carcinogenic activity. NCI1 has multiple molecules and their effectiveness against cellular lung cancer and are labelled positive if they display anti-cancer properties.
Protein graphs (ENZYMES and PROTEINS) are used to predict properties like enzyme-related class labels and taxonomy groups~\citep{borgwardt2005protein}. 
Other examples of commonly used biological datasets belong to the Open Graph Benchmark framework, including obgn-molhiv, obgn-molpcba, obgn-ppa~\citep{hu2020open}.

Social network datasets are employed to classify the networks into specific labels. These networks are constructed through stardom or coauthorship relations.
IMDB-B and IMDB-Multi are actor graphs where nodes represent actors and edges indicate co-starring in a movie. Graph labels correspond to the movie genres, such as romance or action.
 COLLAB is an academic collaboration network comprising egocentric graphs obtained from three public physics-related collaboration datasets. 
Reddit datasets contain graphs of users where edges denote replies between users, and graphs labels are different types of subreddits such as question-answering or discussion-based ones~\citep{IMDB}.

A common feature of the graph classification datasets is that the sizes of the graphs are typically small.
This is often related to the actual domains the networks are obtained from, e.g., molecules with tens of nodes.
Such graphs have limited relational information and hence the datasets they are part of do not serve as true testbeds where the complex graph structure can be utilized for the classification task.
Some recent effort has been attempted to address this issue. MalNet consists of function call graphs where nodes are functions and edges are the calls among them~\citep{freitas2020large}.
However, there are drastically many duplicate function call-graphs in it due to methodological errors in the data collection process.

\subsection{Trend manipulation}

Although our main focus is on graph classification, we also make contributions to the broader area of misinformation and propaganda online by proposing a dataset of coordinated campaigns. Such campaigns aiming to influence public opinion are a common issue in the social media ecosystem. Past studies studied user behavior~\citep{cao2015detecting}, content~\citep{lee2011content,lee2014campaign}, strategies~\citep{zannettou2019disinformation,elmas2023misleading}, and networks to understand and detect coordinated campaigns. Studies focusing on networks investigated the cases of accounts determined to be inauthentic by Twitter~\citep{merhi2023information}, automated accounts (bot)~\citep{minnich2017botwalk,elmas2022characterizing}, follow back accounts~\citep{beers2023followback,elmas2024teamfollowback}, accounts promoting sponsored topics~\citep{varol2017early}, and cryptocurrencies~\citep{tardelli2022detecting}. Additionally there have been instances when GCNs were leveraged to help with tasks like fake news detection~\citep{dou2021user} and rumor detection~\citep{bian2020rumor}. In this study, we present a special case of a network where the users organize themselves to promote a topic as part of their campaign. This has not been studied to date to the best of our knowledge as it is hard to acquire ground truth, i.e., it is not possible to know for which topics the users organized among themselves to promote it as a campaign. 

We provide a ground truth of topics that are coordinated campaigns using fake trends. Trend manipulation has been studied in different contexts. Jakesch et al.~\citeyear{jakesch2021trend} found that political trolls aligned with the Indian ruling party BJP coordinate on WhatsApp groups to mention hashtags in a coordinated manner to make them trending. They reported 75 hashtag manipulation campaigns. Kausar et al.~\citeyear{kausar2021towards} detected the bots and showed that bots are more likely to manipulate political trending topics in Pakistan. 

Our work distinguishes itself by providing the first large-scale annotated dataset of fake Twitter trends for which we have hard proof that bots were used to push them to the trends list. We extend the work of Elmas et al. by reformulating the classification of fake Twitter trends as a graph classification problem~\citep{elmas2021ephemeral,elmas2023analyzing}.

\section{Engagement networks: campaign or not}
\label{sec:Dataset_Collection}
We collect two types of data: campaigns and non-campaigns. We collect campaigns by detecting ephemeral astroturfing attacks in real-time. We collect non-campaigns by manually annotating the popular Twitter trends that were not targeted by the ephemeral astroturfing attacks. We now describe each data collection methodology in detail.

\begin{figure}[!t]
    \centering
    \begin{subfigure} 
        \centering
        \includegraphics[width=0.2\textwidth]{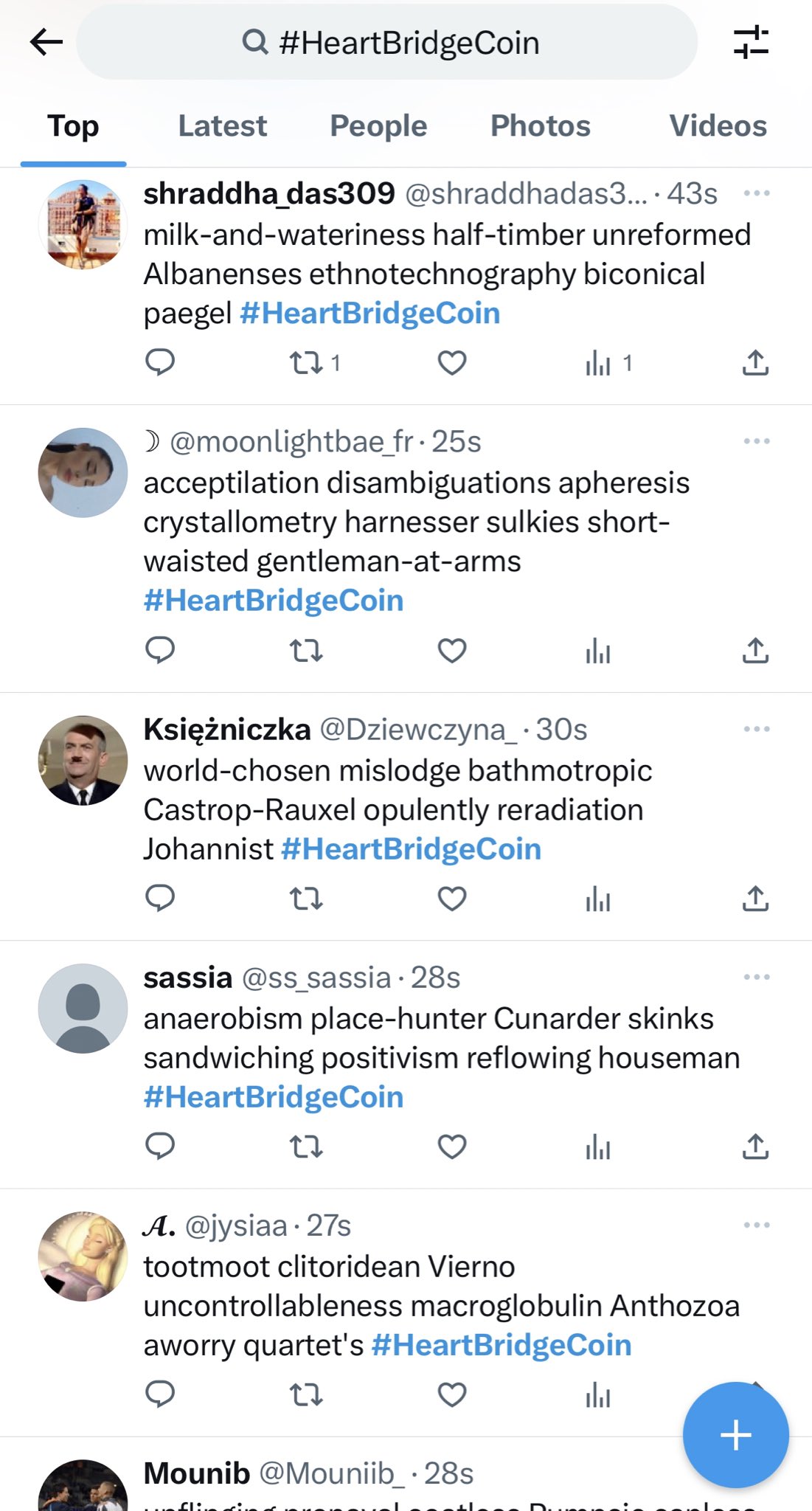} 
        \label{fig:subfig1}
    \end{subfigure}
    \hfill 
    \begin{subfigure} 
        \centering
        \includegraphics[width=0.22\textwidth]{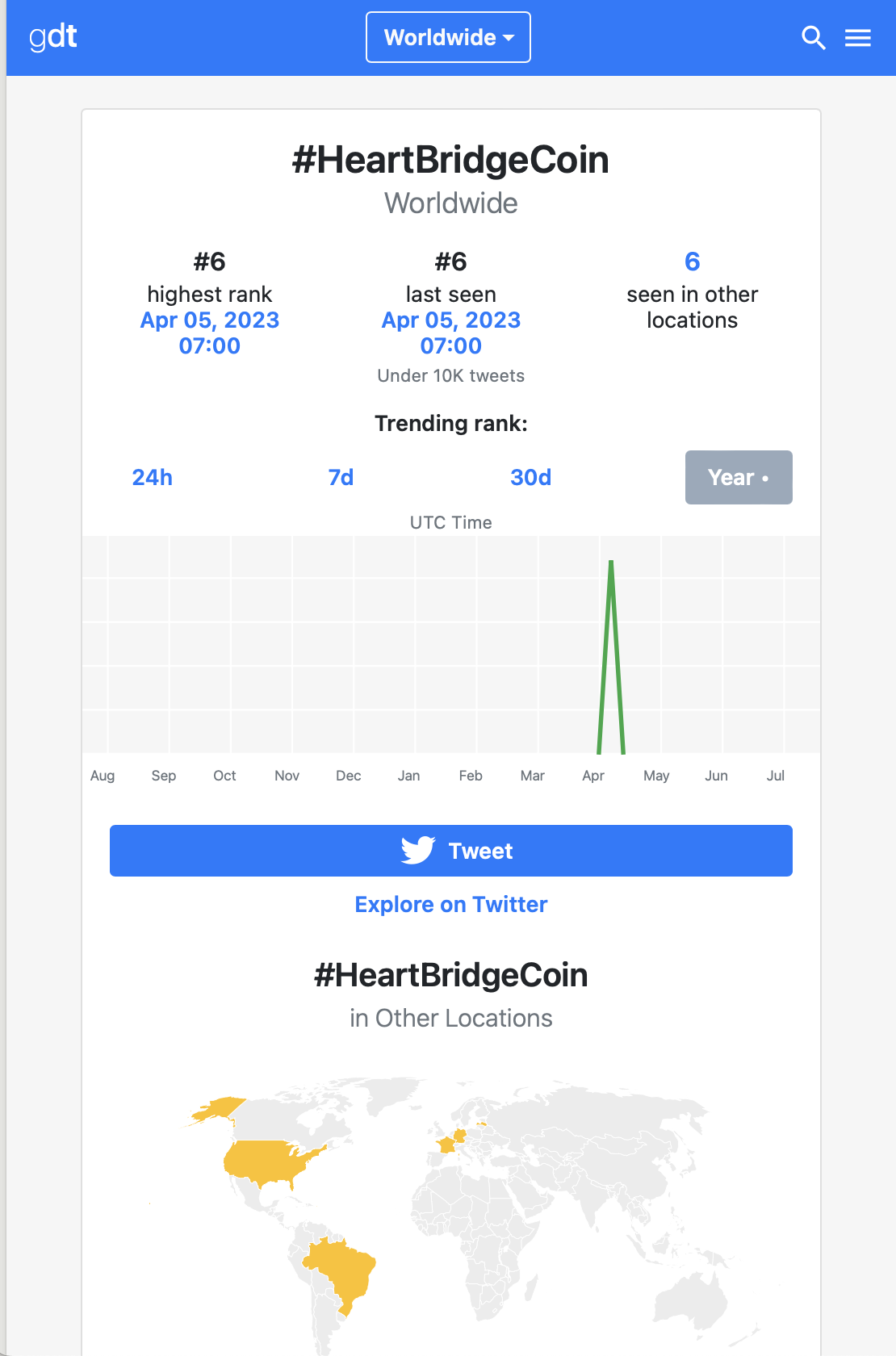} 
        \label{fig:subfig2}
    \end{subfigure}
    \caption{(Left) Randomly generated (lexicon) tweets from bots promoting the hashtag \#HeartBridgeCoin. (Right) It becomes trending in 6 countries and globally for the first and the last time.}
    \label{fig:both}
    \vspace{-2ex}
\end{figure}



\begin{table*}[htbp]
    \centering
    \fontsize{9pt}{11pt}\selectfont
    \setlength{\tabcolsep}{1.5pt}
    \begin{tabular}{l|l|r|r|r|r|r|r|r|l}
    & \fontsize{10}{12}\selectfont \textbf{Sub-types} 
    & \fontsize{10}{12}\selectfont \textbf{\# G} 
    & \multicolumn{3}{|c|}{\fontsize{10}{12}\selectfont \textbf{\# nodes}} 
    & \multicolumn{3}{|c|}{\fontsize{10}{12}\selectfont \textbf{\# edges}} 
    & \fontsize{10}{12}\selectfont \textbf{Explanation}\\

    \hline
         & & & \fontsize{10}{12}\selectfont \textbf{Min} & \fontsize{10}{12}\selectfont \textbf{Max} & \fontsize{10}{12}\selectfont \textbf{Avg} & \fontsize{10}{12}\selectfont \textbf{Min} & \fontsize{10}{12}\selectfont \textbf{Max} & \fontsize{10}{12}\selectfont \textbf{Avg} & \\
    \hline
        \multirow{8}{*}{\rotatebox[origin=c]{90}{Campaign}} & Politics & 62 & 100 & 50,286 & 6,570 & 203 & 71,704 & 10,210 & {\fontsize{9}{11}\selectfont Political promotions, slogans, misinformation camp.}\\ 
                                            & Reform & 58 & 131 & 19,578 & 1,229 & 540 & 1,105,918 & 25,268 & {\fontsize{9}{11}\selectfont People organized for political reforms.}\\ 
                                            & News & 24 & 581 & 54,996 & 10,368 & 942 & 80,784 & 15,582 & {\fontsize{9}{11}\selectfont News pumped up by bots and trolls for more attention.}\\ 
                                            & Finance & 14 & 273 & 9,976 & 1,802 & 243 & 10,725 & 2,334 & {\fontsize{9}{11}\selectfont Finance marketing (mostly cryptocurrency).}\\ 
                                            & Noise & 9 & 454 & 55,933 & 12,180 & 473 & 48,937 & 10,882 & {\fontsize{9}{11}\selectfont Cannot be put in any type.}\\ 
                                            & Cult & 6 & 313 & 7,880 & 2,303 & 637 & 11,615 & 3,431 & {\fontsize{9}{11}\selectfont Slogans by a famous cult with immense access to bots.}\\ 
                                            & Entertainment & 3 & 678 & 4,220 & 2,237 & 3,806 & 132,013 & 48,767 & {\fontsize{9}{11}\selectfont Celebrities attempting to promote themselves.}\\ 
                                            & Common & 3 & 3,487 & 9,974 & 5,919 & 2,818 & 9,470 & 7,066 & {\fontsize{9}{11}\selectfont Common sub-strings combined without known reasons.}\\ 
         & {\bf Overall} & {\bf 179} & {\bf 100} & {\bf 55,933} & {\bf 5,157} & {\bf 203} & {\bf 1,105,918} & {\bf 16,006} &  \\
    \hline
        \multirow{9}{*}{\rotatebox[origin=c]{90}{Non-Campaign}} & News & 52 & 818 & 95,575 & 24,834 & 709 & 213,444 & 43,201 & {\fontsize{9}{11}\selectfont Popular events, sourced outside Twitter.}\\ 
                                                & Sports & 30 & 469 & 75,653 & 9,530 & 403 & 101,656 & 12,948 & {\fontsize{9}{11}\selectfont Popular sports events.}\\ 
                                                & Festival & 17 & 885 & 119,952 & 35,466 & 803 & 199,305 & 55,947 & {\fontsize{9}{11}\selectfont About festivals, holidays, special days}.\\ 
                                                & Internal & 11 & 4,188 & 87,720 & 33,061 & 4,374 & 196,103 & 54,442 & {\fontsize{9}{11}\selectfont Popular events, sourced inside Twitter}.\\ 
                                                & Common & 10 & 1,214 & 64,320 & 17,079 & 1,270 & 99,306 & 24,869 & {\fontsize{9}{11}\selectfont Common substrings combined by people.}\\ 
                                                & Entertainment & 8 & 1,477 & 20,060 & 7,289 & 1,712 & 45,211 & 12,578 & {\fontsize{9}{11}\selectfont Popular TV shows and Youtube videos.}\\ 
                                                & Announ. cam. & 4 & 6,650 & 26,358 & 13,382 & 14,362 & 50,864 & 24,817 & {\fontsize{9}{11}\selectfont Official campaigns launched by major political parties.}\\ 
                                                & Sports cam. & 3 & 2,880 & 4,661 & 3,654 & 4,451 & 7,367 & 5,534 & {\fontsize{9}{11}\selectfont Hashtags launched by popular sports teams.}\\ 
     & {\bf Overall} & {\bf 135} & {\bf 469} & {\bf 119,952} & {\bf 20,632} & {\bf 403} & {\bf 213,444} & {\bf 33,765} &  \\
    \end{tabular}
    \caption{Statistics of the engagement networks for LEN which has 314 networks. }
    \label{tab:dataset_properties}    
\end{table*}

\subsection{Campaigns collection methodology}
\label{subsec:camp_collec}
Adversaries utilize a sophisticated attack named ``Ephemeral Astroturfing'' to generate Twitter trends from scratch. It works in the following way: First, the adversaries select a target keyword to push to trends. This is often motivated by a commercial exchange, i.e. an individual or a group sponsors the attack so that their slogan becomes visible to a wider audience through trends. The adversaries deploy hundreds or thousands of bots to mention this keyword in a coordinated manner. To bypass Twitter's spam filters, they generate tweets by randomly picking up words from a lexicon. These tweets are immediately deleted after being posted. Twitter's trending algorithm does not take the deletions into account and marks the target keywords as trending, which is a security vulnerability. Once the target keyword becomes trending, other users, typically affiliated with the trend sponsors who know about the attack, begin mentioning it to further amplify the visibility of it and their messages. Twitter acknowledged this issue but has not mitigated it~\citep{elmas2021ephemeral}. These attacks are commonly employed in Turkey for political manipulation and advertising purposes. They have also been observed in Brazil and the United States on a few occasions~\citep{elmas2023analyzing}. Figure \ref{fig:both} illustrates an example hashtag promoted through lexicon-generated tweets in English, trending across multiple countries.

To detect the fake trends created by this attack, we used the same methodology described in ~\citep{elmas2021ephemeral, elmas2023analyzing}. We collected the 1\% sample of all tweets posted in real-time using Twitter API. We limited our focus to Turkey where this attack is the most prevalent and only collected Turkish tweets. We used a rule-based classifier to detect tweets that are randomly generated using a Turkish lexicon. The classifier marks a tweet as a lexicon tweet if it is made up of 2-9 tokens, has no punctuation, and begins with a lowercase, which is an anomalous pattern. 4 consecutive lexicon tweets mentioning the same hashtag or a unigram in the sample that are later deleted signify that the hashtag is being promoted by an ephemeral attack.

This would be roughly 400 tweets with the same hashtag and text pattern posted within seconds if we had access to 100\% of Twitter data. While straightforward, this methodology is proven effective in detecting the fake trends created using this attack, scoring 100\% precision and 99\% recall previously~\citep{elmas2021ephemeral}. In this dataset, we observed only two false positives - ``one'' and ``May'' - which we addressed by discarding target keywords with less than five characters.

Between March and May 2023, prior to the Turkish general elections on May 14, 2023, which were marked by intense political campaigning, we identified 190 instances of fake Twitter trends. Subsequently, in July 2023, we conducted a comprehensive collection of all tweets referencing these fake trends within a two-day period. Crucially, by this time, the tweets generated by astroturfing bots had been removed, allowing us to mitigate the noise they typically generate. It is important to note that these bots were not integral to the campaign, but rather employed solely to fabricate fake trends. We removed 20 trends for which we had less than 1000 posts by this time. Those trends may not be strongly backed up by a coordinated campaign. Alternatively, Twitter may have purged their tweets. We annotated the remaining 179 trends as campaigns.


We examined and manually annotated the trends according to the type of campaign they promote, using the labels in~\citep{elmas2021ephemeral}. Annotations are performed by two Turkish-speaking researchers, and conflicts between those two are handled by a third researcher.
Table \ref{tab:dataset_properties} shows the campaign types and descriptions. Out of the 179 trends, 24 were associated with news items that may have sparked genuine discussion among social media users. However, adversaries used bots to further amplify them which may be due to political purposes. For instance, when a politician left his party and criticized it, the rival parties amplified his name as part of their campaign. For 9 campaign trends, we could not ascertain a specific group promoting a campaign related to the topic. Despite this uncertainty, we retained these trends in our analysis, labeling them as ``noise.''.

\begin{table}[!t]
    \setlength{\tabcolsep}{1mm} 
    \fontsize{9pt}{11pt}\selectfont
    \centering
    \begin{tabular}{l|l|r|r|r|r|r|r|r}
    & \fontsize{10}{12}\selectfont \textbf{Sub-type} & \fontsize{10}{12}\selectfont \textbf{\# G}  & \multicolumn{3}{|c|}{\fontsize{10}{12}\selectfont \textbf{\# nodes}} & \multicolumn{3}{|c}{\fontsize{10}{12}\selectfont \textbf{\# edges}}\\
    \hline
    & & & \fontsize{10}{12}\selectfont \textbf{Min} & \fontsize{10}{12}\selectfont \textbf{Max} & \fontsize{10}{12}\selectfont \textbf{Avg} & \fontsize{10}{12}\selectfont \textbf{Min} & \fontsize{10}{12}\selectfont \textbf{Max} & \fontsize{10}{12}\selectfont \textbf{Avg} \\

    \hline
     \multirow{6}{*}{\rotatebox[origin=c]{90}{Campaign}} &       Politics        & 14 & 100 & 1,908 & 805 & 203 & 2,000 & 1108 \\
    &        Reform          & 16 & 131 & 634 & 297 & 540 & 2,027 & 1192 \\
    &        News            & 3 & 581 & 1,671 & 1123 & 942 & 1,726 & 1410 \\
    &        Finance         & 9 & 273 & 1,590 & 775 & 243 & 1,862 & 1024 \\
    &        Noise           & 5 & 454 & 2,520 & 1060 & 473 & 1,634 & 1074 \\
    &                Cult            & 4 & 313 & 705 & 512 & 637 & 1,035 & 843 \\
    & \textbf{Overall} & \textbf{51} & \textbf{100}  & \textbf{2,520} & \textbf{661}  & \textbf{203} & \textbf{2,027} & \textbf{1113} \\     \hline
    \multirow{7}{*}{\rotatebox[origin=c]{90}{Non-Campaign}} &                News            & 10 & 818 & 6,169 & 3757 & 709 & 9,076 & 4578 \\
    &                Sports          & 23 & 469 & 8,355 & 3357 & 403 & 9,998 & 3994 \\
    &                Festival        & 2 & 885 & 5,982 & 3433 & 803 & 6,509 & 3656 \\
    &                Internal        & 1 & 4,188 & 4,188 & 4,188 & 4,374 & 4,374 & 4374 \\
    &                Common          & 5 & 1,214 & 4,962 & 2,989 & 1,270 & 6,277 & 3559 \\
    &                Enter.   & 5 & 1,477 & 7,739 & 4,391 & 1,712 & 10,608 & 6021 \\
    &                Sp. cam.     & 3 & 2,880 & 4,661 & 3,654 & 4,451 & 7,367 & 5534 \\
    &                \textbf{Overall} & \textbf{49} & \textbf{469}  & \textbf{8,355} & \textbf{3545} & \textbf{403} & \textbf{10,608} & \textbf{4364} 
    \end{tabular}
    \caption{Statistics of the engagement networks for the small dataset with 100 networks. This is simply the smallest 100 networks, out of 314, with respect to node counts.}
    \label{tab:small_dataset_properties}
\end{table}

\subsection{Non-campaigns collection methodology}
\label{subsec:non_camp_collec}
We acquired the ground truth for the campaigns by detecting bot activity that specifically aims at trending topics. However, we cannot assume that trends that do not observe such activity are devoid of coordinated efforts since other types of activities (e.g., organizing through messaging apps) may still be the main drivers. Thus, we do a round of manual annotation of the trends that are not classified as part of an ephemeral astroturfing activity. We make the following assumption: the trends associated with external events that attract nationwide interest are more likely to be organic, as their popularity is more likely driven by people tweeting independently, rather than by coordinated efforts. Alternatively, adversaries would be less inclined to campaign using topics that already trending due to external events, as their messages risk being overshadowed by organic discourse. We annotated the trends between March and May 2023 that are 1) person or location names due to a news related to them (49); 2) news that are originally sourced from internal discussions but later made to the mainstream media and became external events (11); 3) popular sports (mostly football) events (30), TV or YouTube shows (8); 4) special days (17); and 5) common hashtags (e.g., \#NewProfilePic) or unigrams (10). 7 hashtags signify a campaign (announced political or sports campaigns), but those hashtags and their campaigns were discussed widely. We discarded the trends that did not fit those categories. We annotated 135 non-campaigns in total. The annotation is not exhaustive but done conservatively to maximize precision.

\begin{table*}[!t]
\normalsize
    \normalsize
    \centering
    \begin{tabular}{l|l|l|l|l|l}
         & \fontsize{10}{12}\selectfont \textbf{Model} & \fontsize{10}{12}\selectfont \textbf{Accuracy} & \fontsize{10}{12}\selectfont \textbf{Precision} & \fontsize{10}{12}\selectfont \textbf{Recall} & \fontsize{10}{12}\selectfont \textbf{F1-Score}\\ 
        \hline

        \multirow{6}{*}{\rotatebox[origin=c]{90}{LEN-small}} 
                                    & Text + MLP & $0.715 \pm 0.019$ & $0.705 \pm 0.011$ & $0.738 \pm 0.038$ & $0.721 \pm 0.024$ \\
                                    & GCN & $0.832 \pm 0.078$ & $0.909 \pm 0.138$ & $0.750 \pm 0.000$ & $0.816 \pm 0.064$\\
                                    & GAT  & $0.856 \pm 0.048$ & $0.871 \pm 0.090$ & $\textbf{0.833} \pm \textbf{0.000}$ & $0.850 \pm 0.043$\\
                                    & GIN & $0.840 \pm 0.000$ & $\textbf{1.000} \pm \textbf{0.000}$ & $0.667 \pm 0.000$ & $0.800 \pm 0.000$\\
                                    & GraphSAGE & $\textbf{0.900} \pm \textbf{0.033}$ & $0.964 \pm 0.073$ & $0.818 \pm 0.000$ & $0.884 \pm 0.033$ \\
                                    & GINE & $0.800 \pm 0.160$ & $0.896 \pm 0.208$ & $0.800 \pm 0.100$ & $0.815 \pm 0.083$\\
                                    & VNGE & $0.875 \pm 0.000$ & $0.877 \pm 0.000$ & $0.818 \pm 0.000$ & $\textbf{0.857} \pm \textbf{0.000}$ \\
                                    & LSD & $0.833 \pm 0.000$ & $0.833 \pm 0.000$ & $0.818 \pm 0.000$ & $0.818 \pm 0.000$ \\
        \hline
        \multirow{6}{*}{\rotatebox[origin=c]{90}{LEN}} 
                                            & Text + MLP & $0.57 \pm 0.018$ & $0.581 \pm 0.02$ & $0.891 \pm 0.067$ & $0.701 \pm 0.012$ \\
                                            & GCN & $0.702 \pm 0.018$ & $0.869 \pm 0.030$ & $0.570 \pm 0.025$ & $0.687 \pm 0.021$\\
                                            & GAT  & $0.735 \pm 0.015$ & $0.783 \pm 0.032$ & $0.752 \pm 0.056$ & $0.765 \pm 0.018$\\
                                            &  GIN & $0.633 \pm 0.065$ & $0.676 \pm 0.091$ & $0.791 \pm 0.157$ & $0.710 \pm 0.037$\\
                                            & GraphSAGE & $0.729 \pm 0.006$ & $\textbf{0.930} \pm \textbf{0.001}$ & $0.578 \pm 0.011$ & $0.713 \pm 0.008$ \\
                                            & GINE & $0.648 \pm 0.091$ & $0.673 \pm 0.121$ & $\textbf{0.896} \pm \textbf{0.139}$ & $0.748 \pm 0.035$\\
                                            & VNGE & $\textbf{0.747} \pm \textbf{0.000}$ & $0.759 \pm 0.000$ & $0.717 \pm 0.000$ & $0.767 \pm 0.000$ \\
                                            & LSD & $0.734 \pm 0.000$ & $0.734 \pm 0.000$ & $0.848 \pm 0.000$ & $\textbf{0.788} \pm \textbf{0.000}$ \\

    \end{tabular}
    \caption{Campaign vs. non-campaign classification. Text + MLP is the non-graph based classifier. The best results are in \textbf{bold}.}
    \label{tab:bin_classification}
\end{table*}

\subsection{Building networks}
\label{subsec:build_graph}
Using the data collected in the last two sections, we build engagement networks. The nodes in the networks represent the users on Twitter and a directed edge from a node A to node B signify that A engaged with (retweeted, replied to, or quoted) B. Some users engaged with the same user repeated times. We only consider their latest engagement. In this process, we retain around 74\% of edges across all the networks.

We use profile and tweet data to assign the attributes of nodes and edges respectively. We used the user description (bio), follower count, following count, user's total tweet count, and user's verification status as node attributes. The edge attributes are features of the tweets that are the user engaged with: the type of engagement (retweet, reply, or quote), text, impression count, engagement count (e.g., number of retweets), number of likes, the timestamp of the tweet and whether the tweet is labeled as sensitive or not. The author's description and the text of the tweet are encoded using an established text encoder called LaBSE~\citep{feng2020language}. The LaBSE model is an bidirectional encoder, that takes source and target translation pairs and embeds them into the same space. The text encoder is initialized with a pre-trained masked language model (MLM) and a translation language model (TLM), which are then concatenated to produce a text embedding. The model is trained using trained using in-batch negative sampling. For our work, we used the pre-trained set of weights for the LaBSE encoder.

LEN comprises of 314 graphs of which 179 are campaign and 135 are non-campaign. 
Table \ref{tab:dataset_properties} presents important statistics.
There are 7 sub-types in campaign and 8 in non-campaign.
Overall, the number of nodes vary between 100 and 119,952 with an average of 11,769, and number of edges are in the range of 203 and 1,105,918 with a mean of 23,593.

To facilitate fast experiments, we also create a smaller, balanced, version of LEN, named LEN-small, that includes 100 networks of the smallest size in LEN. LEN-small consists of 51 campaign and 49 non-campaign networks, details are shown in Table \ref{tab:small_dataset_properties}.
Note that the largest connected component in campaign and non-campaign graphs contain around $76\%$ and $81\%$ of the nodes on average, respectively. Such statistics are provided in Appendix (Table \ref{tab:connected}).

In LEN-small, the number of nodes vary between 100 and 8,355 with an average of 2,079 and number of edges are in the range of 203 and 10,608 with a mean of 2,696.


\section{Graph classification on engagement networks}
\label{sec:exps}






To understand the challenges of classifying networks in LEN, we experiment with several established graph classification methods to perform binary classification, campaign vs. non-campaign, and multi-class classification, which is classifying the type of campaign.


\textbf{Experimental setup:} 
For all of our experiments, we utilize stratified random sampling to split the data into 75\% training and 25\% testing sets.
For binary classification (campaign vs non-campaign), we measure model performance using accuracy, precision, recall, and F1-Score. 
For multi-class classification (campaign type), we use accuracy, weighted precision/recall, and micro/macro F1-Scores. 
The experiments were conducted on a Linux operating system (v. 3.10.0-1127) running on a machine with Intel(R) Xeon(R) Gold 6130 CPU processor at 2.10 GHz with 192 GB memory. 
An Nvidia A100 GPU was used specifically for the GNN experiments.
{\bf Our code is publicly available at \url{https://github.com/erdemUB/LEN}.}

\textbf{Non-graph based classifier:}
To emphasize the impact of graph structure, we use a non-graph based classifier that uses the user description and tweets in an engagement network along with a MLP for downstream classification tasks. For this, we use the mean user caption embedding and mean tweet embedding, both of which can be obtained by averaging the user caption embeddings for all users or tweets in the engagement network. The mean user caption embedding and tweet embedding are concatenated and passed through an MLP. The user captions and tweets are encoded using the Conditional Masked Language Modeling.

\textbf{Graph classifiers:}
We use five established Graph Neural Network (GNN) architectures for evaluation. 

    (1) Graph Convolutional Network (GCN): Leverages a technique called ``neural message passing'' to learn node representations~\citep{kipf2016semi}. A node's embedding is updated by aggregating and combining the embeddings of its neighboring nodes. These neighborhood embeddings are normalized using the diagonal degree matrix to account for the varying number of neighbors each node may have.
    (2) Graph Attention Network (GAT):
    Also employs a message-passing approach to learn node representations~\citep{velivckovic2018graph}. Different from GCN, GAT incorporates an attention mechanism during message aggregation which assigns weights to incoming messages from neighboring nodes, focusing the node's representation on the most informative neighbors.
    (3) Graph Isomorphism Network (GIN):
    A provably more-expressive GNN which is as  powerful as the Weisfeiler-Lehman test in distinguishing isomorphic graphs~\citep{xu2018powerful}. The architecture aggregates neighborhood embeddings similar to GCN's but additionally passes it through a MLP, after each layer, to make the architecture more expressive. Additionally, GIN also weights out the importance of the ego node using a parameter $\epsilon$ where a high value gives more importance to the node compared to its neighbors.
    (4) GraphSAGE: Provides an inductive representational learning capability, thanks to its ability to generalize to unseen nodes, unlike transductive models~\citep{hamilton2017inductive}. This is done by learning a message-passing model on a sampled set of nodes in the given graph.
    (5) Edge attribute GIN (GINE): To leverage the additional information present in edge features, we use a modified version of the GIN architecture, called GINE. Here the node features of the neighboring nodes along with the edge features are added along the respective edges, before aggregating them in the message-passing function. 
    
Additionally, we use two non-neural network based graph embedding models, namely VNGE and LSD. (1) VNGE: Approximates the spectral distances between graphs using the Von Neumann Graph Entropy (VNGE) by measuring information divergence/distance between graphs~\citep{chen2019fast}.
(2) NetLSD: Measures the spectral distance between graphs using the heat kernel~\citep{tsitsulin2018netlsd}. Both models are approximated using SLaQ, which helps approximate spectral distances. To do this, SLaQ takes in two parameters, namely, number of random steps ($n_v$) and number of Lanczos steps ($s$).

These GNNs can handle both directed and undirected graphs, allowing us to directly apply them to our directed networks without modification. 
Initially, each graph is processed by a 2-layer GNN to generate informative node embeddings and those are combined using global mean pooling to create a single graph-level embedding. 
Lastly, we utilize a two-layer MLP to predict the class.

To demonstrate the difficulty of classifying large engagement networks, we perform several experiments.

We use the established GNNs as graph classifiers, described before.
We conduct three experiments: (1) Binary classification to distinguish campaign networks from non-campaign networks; (2) Multi-class classification to categorize campaigns based into the 7 sub-types as shown in Table 2; and (3) Binary classification to identify if a trending topic signifying news is a campaign or not. 
We ensure a fair comparison across the four GNN architectures by tuning hyperparameters: $l\in$ $\{0.001, 0.0001, 0.00001\}$ and hidden layer dimension h $\in$ $\{128, 256, 512, 1024\}$. For each combination, we ran our model five times with different random seeds and report the average scores. Similarly, for VNGE and LSD, we tune the models by trying all combinations of $n_v \in \{10, 15, 20\}$ and $s \in \{10, 15, 20\}$.

\begin{table*}[!t]
    \centering
    \normalsize
        \begin{tabular}{l|l|l|l|l|l|l}
            & \fontsize{10}{12}\selectfont \textbf{Model} & \fontsize{10}{12}\selectfont \textbf{Accuracy} & \fontsize{10}{12}\selectfont \textbf{Precision} & \fontsize{10}{12}\selectfont \textbf{Recall} & \fontsize{10}{12}\selectfont \textbf{Micro F1} 
            & \fontsize{10}{12}\selectfont \textbf{Macro F1}\\ 
\hline

                    \multirow{7}{*}{\rotatebox[origin=c]{90}{LEN-small}} 
                                                    & Text + MLP & $0.367 \pm 0.041$ & $0.209 \pm 0.135$ & $0.367 \pm 0.041$ & $0.367 \pm 0.041$ & $0.133 \pm 0.041$\\

                                                    & GCN & $0.533 \pm 0.041$ & $0.371 \pm 0.042$ &       $0.533 \pm 0.041$ & $0.533 \pm 0.041$ & $0.251 \pm 0.022$\\
	                                               & GAT & $0.567 \pm 0.033$ & $0.387 \pm 0.031$ & $0.567 \pm 0.033$ & $0.567 \pm 0.033$ & $0.264 \pm 0.014$\\
                                                  & GIN & $0.633 \pm 0.067$ & $0.484 \pm 0.105$ & $0.633 \pm 0.067$ & $0.633 \pm 0.067$ & $0.351 \pm 0.091$\\
                                                  & GraphSAGE & $0.583 \pm 0.053$ & $0.470 \pm 0.082$ & $0.583 \pm 0.053$ & $0.583 \pm 0.053$ & $0.320 \pm 0.061$ \\
                                                  & GINE & $0.650 \pm 0.033$ & $0.569 \pm 0.040$ & $0.650 \pm 0.033$ & $0.650 \pm 0.033$ & $0.361 \pm 0.042$ \\
                                                  & VNGE & $\textbf{0.833} \pm \textbf{0.000}$ & $\textbf{0.771} \pm \textbf{0.000}$ & $\textbf{0.833} \pm \textbf{0.000}$ & $\textbf{0.833} \pm \textbf{0.000}$ & $\textbf{0.671} \pm \textbf{0.000}$ \\
                                                  & LSD  & $0.667 \pm 0.000$ & $0.594 \pm 0.000$ & $0.667 \pm 0.000$ & $0.667 \pm 0.000$ & $0.414 \pm 0.000$ \\
        \hline
                    \multirow{7}{*}{\rotatebox[origin=c]{90}{LEN}} & GCN & $0.641 \pm 0.009$ & $0.457 \pm 0.008$ & $0.641 \pm 0.009$ & $0.641 \pm 0.009$ & $0.252 \pm 0.004$\\
                                                    & Text + MLP & $0.645 \pm 0.011$ & $0.462 \pm 0.017$ & $0.645 \pm 0.011$ & $0.645 \pm 0.011$ & $0.218 \pm 0.006$\\
       	                                            & GAT  & $0.636 \pm 0.000$ & $0.467 \pm 0.006$ & $0.636 \pm 0.000$ & $0.636 \pm 0.000$ & $0.257 \pm 0.001$\\
                                                    & GIN & $0.659 \pm 0.000$ & $0.495 \pm 0.010$ & $0.659 \pm 0.000$ & $0.659 \pm 0.000$ & $0.269 \pm 0.002$\\
                                                    & GraphSAGE & $0.641 \pm 0.017$ & $0.453 \pm 0.010$ & $0.641 \pm 0.017$ & $0.641 \pm 0.017$ & $0.252 \pm 0.006$ \\
                                                    & GINE & $\textbf{0.677} \pm \textbf{0.009}$ & $0.478 \pm 0.013$ & $\textbf{0.677} \pm \textbf{0.009}$ & $\textbf{0.677} \pm \textbf{0.009}$ & $0.233 \pm 0.004$\\
                                                    & VNGE& $0.659 \pm 0.000$ & $\textbf{0.640} \pm \textbf{0.000}$ & $0.659 \pm 0.000$ & $0.659 \pm 0.000$ & $\textbf{0.383} \pm \textbf{0.000}$ \\
                                                    & LSD & $0.545 \pm 0.000$ & $0.512 \pm 0.000$ & $0.545 \pm 0.000$ & $0.545 \pm 0.000$ & $0.252 \pm 0.000$ \\

    \end{tabular}
    \caption{Campaign type classification for 7 labels: politics, reform, news, finance, cult, entertainment, and common, see Table \ref{tab:dataset_properties} for details. Text + MLP refers to the non-graph based classifier. The best results are in \textbf{bold}.}
    \label{tab:multi_classification}
    \vspace{-3ex}
\end{table*}

\subsection{Binary classification}
\label{subsubsec:bin_classification}

\begin{figure}[!t]
    \centering
    \includegraphics[width=0.48\textwidth]{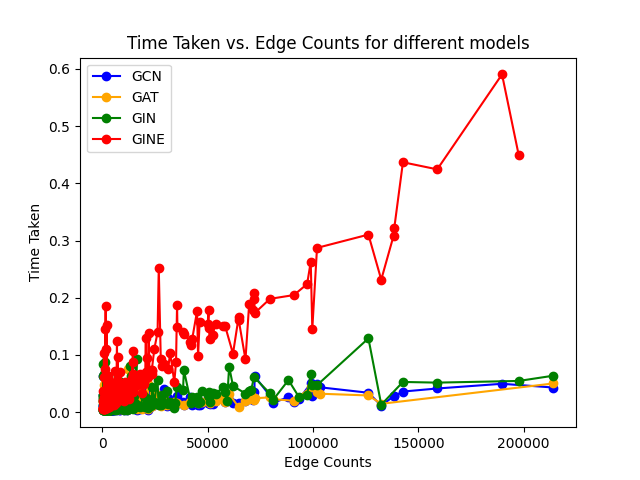}
    \caption{Training runtime (in seconds) vs graph size.}
    \label{fig:timing_plots}
\end{figure}

We first identify campaign networks by distinguishing them from non-campaign networks. Table \ref{tab:bin_classification} summarizes the results for LEN, 179+135 graphs, as well as LEN-small, which has 51+49 networks of smaller size. GraphSAGE and VNGE achieve the best accuracy for the small and the complete dataset, while VNGE and LSD achieve the best F1 scores. ROC curves across different training epochs are given in Appendix (Figure \ref{fig:ROC_small} refers to the ROC curves for the small dataset and Figure \ref{fig:ROC_all} refers to the ROC curves for the complete dataset). One interesting observation is that the accuracy and F1 scores are lower for LEN, which has larger networks than the LEN-small. This highlights the difficulty in classifying large networks, which is expected as most datasets in the graph classification literature contain small networks, as discussed in the related work\ref{sec:relwork}.
Regarding the runtime performance, Figure \ref{fig:timing_plots} presents the time taken to run the graph classification models plotted along the size of the graph. We observe that GCN, GIN and GAT have minor changes in performance with graph size. However, GINE shows a linear growth with the size of the graph.

\begin{table*}[!t]
    \centering
    \begin{tabular}{l|l|l|l|l}
        \fontsize{10}{12}\selectfont \textbf{Model} & \fontsize{10}{12}\selectfont \textbf{Accuracy} 
        & \fontsize{10}{12}\selectfont \textbf{Precision} & \fontsize{10}{12}\selectfont \textbf{Recall} & \fontsize{10}{12}\selectfont \textbf{F1 Score}\\ 
\hline

        Text + MLP& $0.585 \pm 0.062$ & $0.535 \pm 0.040$ & $0.843 \pm 0.000$ & $0.651 \pm 0.031$ \\
        GCN& $0.554 \pm 0.031$ & $0.551 \pm 0.018$ & $\textbf{0.943} \pm \textbf{0.114}$ & $0.692 \pm 0.037$ \\
        GAT & $0.585 \pm 0.092$ & $0.631 \pm 0.185$ & $0.914 \pm 0.171$ & $0.705 \pm 0.011$ \\
        GIN & $0.492 \pm 0.062$ & $0.529 \pm 0.057$ & $0.543 \pm 0.057$ & $0.535 \pm 0.055$ \\
        GraphSAGE & $0.769 \pm 0.028$ & $\textbf{1.000} \pm \textbf{0.000}$ & $0.571 \pm 0.089$ & $0.727 \pm 0.032$\\
        GINE & $0.585 \pm 0.092$ & $0.635 \pm 0.135$ & $0.800 \pm 0.194$ & $0.673 \pm 0.026$ \\
        VNGE & $0.769 \pm 0.020$ & $0.769 \pm 0.024$ & $0.833 \pm 0.049$ & $0.769 \pm 0.063$\\
        LSD & $\textbf{0.769} \pm \textbf{0.001}$ & $0.773 \pm 0.005$ & $0.857 \pm 0.010$ & $\textbf{0.800} \pm \textbf{0.006}$ \\
    \end{tabular}
    \caption{Campaign vs. non-campaign classification for news-based engagement networks. Text + MLP refers to the non-graph based classifier. The best results are in \textbf{bold}.}
    \label{tab:news_classification}
\end{table*}

\begin{figure*}[!t]
    \hspace{-5ex}
    \includegraphics[width=1.07\textwidth]{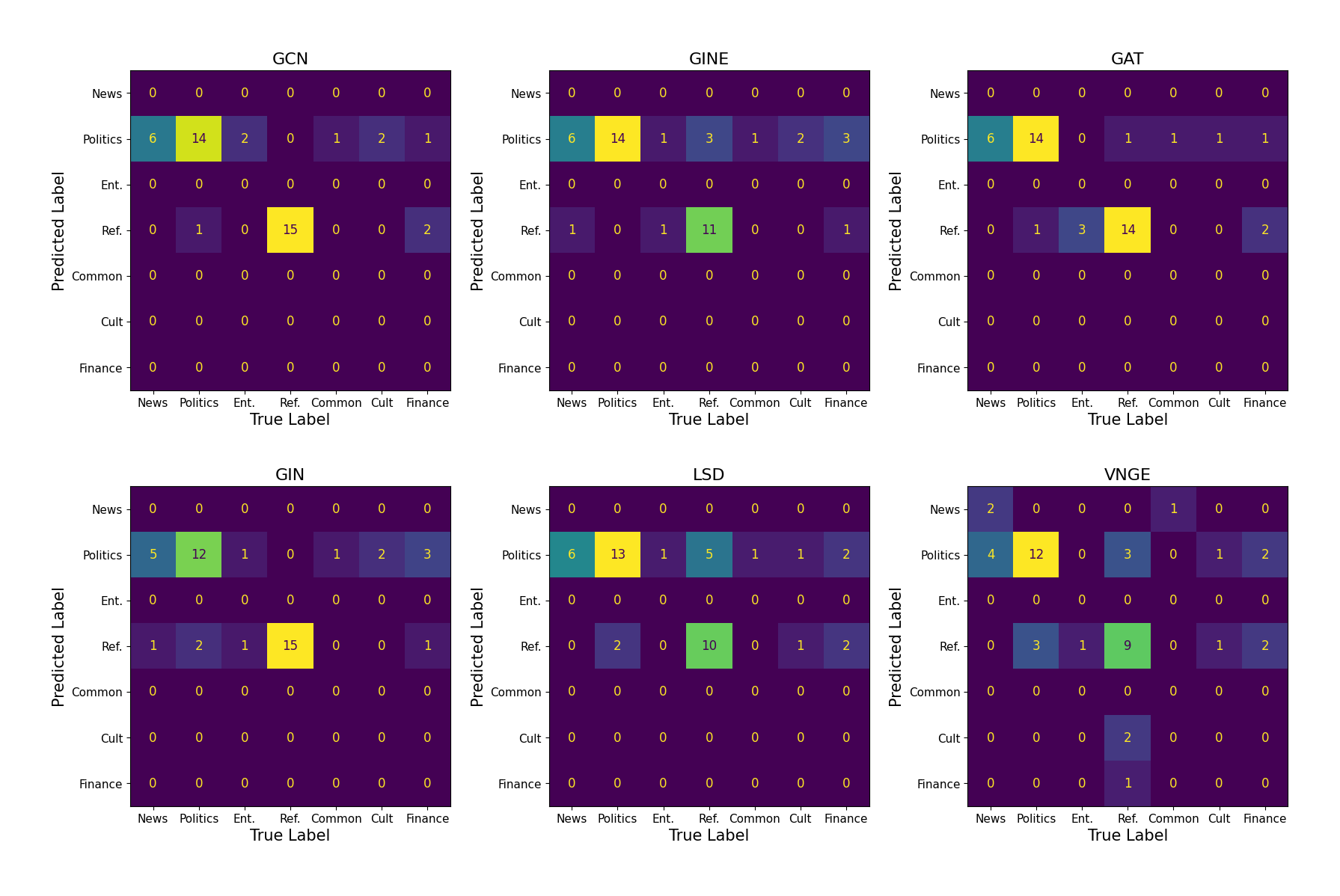}
    \hfill
    \vspace{-7ex}
    \caption{Confusion matrices to display the performance of the graph classifiers.} 
    \label{fig:confusion_matrices}
\end{figure*}


\subsection{Campaign type classification}
\label{subsubsec:multiclass_classification}

We next classify campaign graphs into seven specific types: politics, reform, news, finance, cult, entertainment, and common, as detailed in Table \ref{tab:dataset_properties}.
Identifying the campaigns with potentially negative social impacts (e.g., false political campaigns) by only using the graph structure can be an important problem to understand misinformation. Similar to the binary classification setup above, we use the established GNNs for multi-class classification. Table~\ref{tab:multi_classification} presents the results.

VGNE achieves the highest accuracy in LEN-small and GINE achieves the highest accuracy on LEN. While VGNE and GINE provide high micro F1 scores, we notice that the macro F1 scores are lower. This applies to all the other models. We suspect this is due to imbalanced labels in the data, as shown in Table \ref{tab:dataset_properties}, where some campaign types have significantly fewer graphs. This is further demonstrated by the confusion matrices shown in Figure \ref{fig:confusion_matrices}, where most graphs are classified as either Politics or Reform by the baseline models. Another noteworthy observation is that accuracy and F1 scores for both datasets in multi-class campaign type classification is lower than the scores for binary campaign vs. non-campaign classification (in Table \ref{tab:bin_classification}). This suggests that the task of distinguishing the campaign type is more challenging than simply detecting the campaigns. Overall, the label imbalance within seven classes of large networks is an interesting and challenging direction for graph classification methods and our dataset offers a promising testbed.

\subsection{Campaign vs. non-campaign classification for news-based graphs }
We also investigate a finer-grained binary classification among engagement networks that are based on news.
There are 24 campaign networks within which the news are amplified by bots and trolls, and 52 non-campaign networks that are organically formed due to popular events happening in real world.
We conjecture that this subset is uniquely challenging for classification as they share the same theme but different formation processes.
To address the imbalance, we randomly sample 24 non-campaign graphs and run the GNNs mentioned above using the same setup above. Table \ref{tab:news_classification} gives the results.
LSD performs the best in terms of accuracy and F1 score, similar to the case in binary classification over all networks (Table \ref{tab:bin_classification}).
However, the scores for all the classifiers are consistently lower for the news networks, which again suggests a challenging testbed, especially for the neural network based approaches.

\section{Limitations} \label{sec:limitations}
The most prominent limitation is that the data collection predates the decision to restrict API access~\citep{murtfeldt2024rip}. Twitter revoked access to the API endpoint that provides the 1\% sample of all tweets, rendering real-time detection of bots creating fake trends infeasible as these bots delete their tweets immediately. In addition, collecting large-scale datasets has become prohibitively expensive (\$5000 per month for access to 1M tweets as of May 2024~\citep{twitter_api}). Consequently, it is not possible to reproduce or practically extend this dataset, potentially making it one of the last of its kind.

Another limitation is in our methodology used to build the engagement networks. As part of our experimental setup, while building the graphs, we restrict ourselves to only using the latest interaction between any two users. Although this removes some interactions, we end up preserving about 74\% of the edges across all networks.

We acknowledge a potential bias in our dataset towards popular events, which resulted in larger networks compared to campaign-related events. This bias likely arises because less popular events do not make it to the trends list, and even if they do, they often do not fit our heuristics for annotating non-campaigns and are subsequently excluded from the dataset.

We also acknowledge that adversarial activity on social media is diverse and evolving, and ephemeral astroturfing may not be the only strategy for creating fake Twitter trends. We would like to clarify that we followed Elmas et al.'s~\citeyear{elmas2021ephemeral} findings which suggested that ephemeral astroturfing was the only strategy adversaries employed after 2015 to create fake trends using bots although there were other strategies before. Therefore we assumed it is still the primary strategy while conducting this study. To address the potential issue of misclassifying campaigns created by other malicious strategies as non-campaigns, we manually annotated non-campaigns using our heuristics.

\section{Ethics}
\label{sec:ethics}

Our dataset consists of only users with public profiles. To better protect the privacy of those users, we concealed the identifying information of all users in the public version of our dataset. This aligns with Twitter's policy for sharing information operations accounts, where they publicly share data of malicious accounts but hash the identifying information of those with fewer than 5000 followers~\citep{x2024moderation}. We hashed the following fields: user id, user display name, user screen name (handle), retweeted, mentioned, and replied user id. We would like to clarify that this process does not interfere with developing the baselines employing those datasets. We will grant full access including these fields to the researchers upon reasonable request. 

We would like to clarify that although campaigns in this dataset were supported by bots, and so were inauthentic to some degree, it is unfair to label all of them as fully inauthentic and have absolutely no genuine support. Thus, our work should not be misused to disregard those campaigns as inauthentic or disregard other movements as inauthentic using a classifier trained by this dataset.

\section{Acknowledgements}
A. A. Gopalakrishnan, J. Hossain, and A. E. Sar{\i}y\"{u}ce were supported by NSF awards OAC-2107089 and IIS-2236789, and this research used resources from the Center for Computational Research at the University at Buffalo~\cite{CCR}.

\bibliography{bib}

\begin{thebibliography}{56}
\providecommand{\natexlab}[1]{#1}

\bibitem[{CCR(2025)}]{CCR}
 2025.
\newblock Center for Computational Research, University at Buffalo, \url{http://hdl.handle.net/10477/79221}.

\bibitem[{Alon and Yahav(2020)}]{alon2020bottleneck}
Alon, U.; and Yahav, E. 2020.
\newblock On the bottleneck of graph neural networks and its practical implications.
\newblock \emph{arXiv preprint arXiv:2006.05205}.

\bibitem[{Bacciu, Conte, and Landolfi(2023)}]{bacciu2023graph}
Bacciu, D.; Conte, A.; and Landolfi, F. 2023.
\newblock Graph pooling with maximum-weight k-independent sets.
\newblock In \emph{Thirty-Seventh AAAI Conference on Artificial Intelligence}.

\bibitem[{Beers et~al.(2023)Beers, Schafer, Kennedy, Wack, Spiro, and Starbird}]{beers2023followback}
Beers, A.; Schafer, J.~S.; Kennedy, I.; Wack, M.; Spiro, E.~S.; and Starbird, K. 2023.
\newblock Followback clusters, satellite audiences, and bridge nodes: coengagement networks for the 2020 US election.
\newblock In \emph{Proceedings of the International AAAI Conference on Web and Social Media}, volume~17, 59--71.

\bibitem[{Bian et~al.(2020)Bian, Xiao, Xu, Zhao, Huang, Rong, and Huang}]{bian2020rumor}
Bian, T.; Xiao, X.; Xu, T.; Zhao, P.; Huang, W.; Rong, Y.; and Huang, J. 2020.
\newblock Rumor detection on social media with bi-directional graph convolutional networks.
\newblock In \emph{Proceedings of the AAAI conference on artificial intelligence}, 549--556.

\bibitem[{Bianchi, Grattarola, and Alippi(2020)}]{bianchi2020spectral}
Bianchi, F.~M.; Grattarola, D.; and Alippi, C. 2020.
\newblock Spectral clustering with graph neural networks for graph pooling.
\newblock In \emph{International conference on machine learning}, 874--883. PMLR.

\bibitem[{Bianchi et~al.(2020)Bianchi, Grattarola, Livi, and Alippi}]{bianchi2020hierarchical}
Bianchi, F.~M.; Grattarola, D.; Livi, L.; and Alippi, C. 2020.
\newblock Hierarchical representation learning in graph neural networks with node decimation pooling.
\newblock \emph{IEEE Transactions on Neural Networks and Learning Systems}, 33(5): 2195--2207.

\bibitem[{Borgwardt and Kriegel(2005)}]{borgwardt2005shortest}
Borgwardt, K.~M.; and Kriegel, H.-P. 2005.
\newblock Shortest-path kernels on graphs.
\newblock In \emph{Fifth IEEE international conference on data mining (ICDM'05)}, 8--pp. IEEE.

\bibitem[{Borgwardt et~al.(2005)Borgwardt, Ong, Sch{\"o}nauer, Vishwanathan, Smola, and Kriegel}]{borgwardt2005protein}
Borgwardt, K.~M.; Ong, C.~S.; Sch{\"o}nauer, S.; Vishwanathan, S.; Smola, A.~J.; and Kriegel, H.-P. 2005.
\newblock Protein function prediction via graph kernels.
\newblock \emph{Bioinformatics}, 21(suppl\_1): i47--i56.

\bibitem[{Cao and Caverlee(2015)}]{cao2015detecting}
Cao, C.; and Caverlee, J. 2015.
\newblock Detecting spam urls in social media via behavioral analysis.
\newblock In \emph{Advances in Information Retrieval: 37th European Conference on IR Research, ECIR 2015, Vienna, Austria, March 29-April 2, 2015. Proceedings 37}, 703--714. Springer.

\bibitem[{Center(2024)}]{x2024moderation}
Center, X.~T. 2024.
\newblock Moderation Research.
\newblock Accessed: 2024-05-27.

\bibitem[{Chen et~al.(2019)Chen, Wu, Liu, and Rajapakse}]{chen2019fast}
Chen, P.-Y.; Wu, L.; Liu, S.; and Rajapakse, I. 2019.
\newblock Fast incremental von neumann graph entropy computation: Theory, algorithm, and applications.
\newblock In \emph{International Conference on Machine Learning}, 1091--1101. PMLR.

\bibitem[{Developer(2024)}]{twitter_api}
Developer, T. 2024.
\newblock About the Twitter API.
\newblock Accessed: 2024-05-29.

\bibitem[{Dou et~al.(2021)Dou, Shu, Xia, Yu, and Sun}]{dou2021user}
Dou, Y.; Shu, K.; Xia, C.; Yu, P.~S.; and Sun, L. 2021.
\newblock User preference-aware fake news detection.
\newblock In \emph{Proceedings of the 44th international ACM SIGIR conference on research and development in information retrieval}, 2051--2055.

\bibitem[{Duchenne, Joulin, and Ponce(2011)}]{duchenne2011graph}
Duchenne, O.; Joulin, A.; and Ponce, J. 2011.
\newblock A graph-matching kernel for object categorization.
\newblock In \emph{2011 International conference on computer vision}, 1792--1799. IEEE.

\bibitem[{Elmas(2023)}]{elmas2023analyzing}
Elmas, T. 2023.
\newblock Analyzing activity and suspension patterns of twitter bots attacking turkish twitter trends by a longitudinal dataset.
\newblock In \emph{Companion Proceedings of the ACM Web Conference 2023}, 1404--1412.

\bibitem[{Elmas, Overdorf, and Aberer(2022)}]{elmas2022characterizing}
Elmas, T.; Overdorf, R.; and Aberer, K. 2022.
\newblock Characterizing retweet bots: The case of black market accounts.
\newblock In \emph{Proceedings of the International AAAI Conference on Web and Social Media}, volume~16, 171--182.

\bibitem[{Elmas, Overdorf, and Aberer(2023)}]{elmas2023misleading}
Elmas, T.; Overdorf, R.; and Aberer, K. 2023.
\newblock Misleading repurposing on twitter.
\newblock In \emph{Proceedings of the International AAAI Conference on Web and Social Media}, volume~17, 209--220.

\bibitem[{Elmas et~al.(2021)Elmas, Overdorf, {\"O}zkalay, and Aberer}]{elmas2021ephemeral}
Elmas, T.; Overdorf, R.; {\"O}zkalay, A.~F.; and Aberer, K. 2021.
\newblock Ephemeral astroturfing attacks: The case of fake twitter trends.
\newblock In \emph{2021 IEEE European symposium on security and privacy (EuroS\&P)}, 403--422. IEEE.

\bibitem[{Elmas, Randl, and Attia(2024)}]{elmas2024teamfollowback}
Elmas, T.; Randl, M.; and Attia, Y. 2024.
\newblock \# TeamFollowBack: Detection \& Analysis of Follow Back Accounts on Social Media.
\newblock In \emph{Proceedings of the International AAAI Conference on Web and Social Media}, volume~18, 381--393.

\bibitem[{Feng et~al.(2020)Feng, Yang, Cer, Arivazhagan, and Wang}]{feng2020language}
Feng, F.; Yang, Y.; Cer, D.; Arivazhagan, N.; and Wang, W. 2020.
\newblock Language-agnostic BERT sentence embedding.
\newblock \emph{arXiv preprint arXiv:2007.01852}.

\bibitem[{Freitas et~al.(2020)Freitas, Dong, Neil, and Chau}]{freitas2020large}
Freitas, S.; Dong, Y.; Neil, J.; and Chau, D.~H. 2020.
\newblock A large-scale database for graph representation learning.
\newblock \emph{arXiv preprint arXiv:2011.07682}.

\bibitem[{Frohlich, Wegner, and Zell(2005)}]{frohlich2005assignment}
Frohlich, H.; Wegner, J.~K.; and Zell, A. 2005.
\newblock Assignment kernels for chemical compounds.
\newblock In \emph{Proceedings. 2005 IEEE International Joint Conference on Neural Networks, 2005.}, volume~2, 913--918. IEEE.

\bibitem[{Gebru et~al.(2021)Gebru, Morgenstern, Vecchione, Vaughan, Wallach, Iii, and Crawford}]{gebru2021datasheets}
Gebru, T.; Morgenstern, J.; Vecchione, B.; Vaughan, J.~W.; Wallach, H.; Iii, H.~D.; and Crawford, K. 2021.
\newblock Datasheets for datasets.
\newblock \emph{Communications of the ACM}, 64(12): 86--92.

\bibitem[{Hamilton, Ying, and Leskovec(2017)}]{hamilton2017inductive}
Hamilton, W.; Ying, Z.; and Leskovec, J. 2017.
\newblock Inductive representation learning on large graphs.
\newblock \emph{Advances in neural information processing systems}, 30.

\bibitem[{Hammack et~al.(2011)Hammack, Imrich, Klav{\v{z}}ar, Imrich, and Klav{\v{z}}ar}]{hammack2011handbook}
Hammack, R.~H.; Imrich, W.; Klav{\v{z}}ar, S.; Imrich, W.; and Klav{\v{z}}ar, S. 2011.
\newblock \emph{Handbook of product graphs}, volume~2.
\newblock CRC press Boca Raton.

\bibitem[{Hu et~al.(2020)Hu, Fey, Zitnik, Dong, Ren, Liu, Catasta, and Leskovec}]{hu2020open}
Hu, W.; Fey, M.; Zitnik, M.; Dong, Y.; Ren, H.; Liu, B.; Catasta, M.; and Leskovec, J. 2020.
\newblock Open graph benchmark: Datasets for machine learning on graphs.
\newblock \emph{Advances in neural information processing systems}, 33: 22118--22133.

\bibitem[{Jakesch et~al.(2021)Jakesch, Garimella, Eckles, and Naaman}]{jakesch2021trend}
Jakesch, M.; Garimella, K.; Eckles, D.; and Naaman, M. 2021.
\newblock Trend alert: A cross-platform organization manipulated Twitter trends in the Indian general election.
\newblock \emph{Proceedings of the ACM on Human-computer Interaction}, 5(CSCW2): 1--19.

\bibitem[{Kang, Tong, and Sun(2012)}]{kang2012fast}
Kang, U.; Tong, H.; and Sun, J. 2012.
\newblock Fast random walk graph kernel.
\newblock In \emph{Proceedings of the 2012 SIAM international conference on data mining}, 828--838. SIAM.

\bibitem[{Kausar, Tahir, and Mehmood(2021)}]{kausar2021towards}
Kausar, S.; Tahir, B.; and Mehmood, M.~A. 2021.
\newblock Towards understanding trends manipulation in Pakistan Twitter.
\newblock \emph{arXiv preprint arXiv:2109.14872}.

\bibitem[{Kipf and Welling(2016)}]{kipf2016semi}
Kipf, T.~N.; and Welling, M. 2016.
\newblock Semi-supervised classification with graph convolutional networks.
\newblock \emph{arXiv preprint arXiv:1609.02907}.

\bibitem[{Kriege and Mutzel(2012)}]{kriege2012subgraph}
Kriege, N.; and Mutzel, P. 2012.
\newblock Subgraph matching kernels for attributed graphs.
\newblock \emph{arXiv preprint arXiv:1206.6483}.

\bibitem[{Lee, Lee, and Kang(2019)}]{lee2019self}
Lee, J.; Lee, I.; and Kang, J. 2019.
\newblock Self-attention graph pooling.
\newblock In \emph{International conference on machine learning}, 3734--3743. PMLR.

\bibitem[{Lee, Rossi, and Kong(2018)}]{lee2018graph}
Lee, J.~B.; Rossi, R.; and Kong, X. 2018.
\newblock Graph classification using structural attention.
\newblock In \emph{Proceedings of the 24th ACM SIGKDD International Conference on Knowledge Discovery \& Data Mining}, 1666--1674.

\bibitem[{Lee et~al.(2011)Lee, Caverlee, Cheng, and Sui}]{lee2011content}
Lee, K.; Caverlee, J.; Cheng, Z.; and Sui, D.~Z. 2011.
\newblock Content-driven detection of campaigns in social media.
\newblock In \emph{Proceedings of the 20th ACM international conference on Information and knowledge management}, 551--556.

\bibitem[{Lee et~al.(2014)Lee, Caverlee, Cheng, and Sui}]{lee2014campaign}
Lee, K.; Caverlee, J.; Cheng, Z.; and Sui, D.~Z. 2014.
\newblock Campaign extraction from social media.
\newblock \emph{ACM Transactions on Intelligent Systems and Technology (TIST)}, 5(1): 1--28.

\bibitem[{Merhi, Rajtmajer, and Lee(2023)}]{merhi2023information}
Merhi, M.; Rajtmajer, S.; and Lee, D. 2023.
\newblock Information operations in turkey: Manufacturing resilience with free twitter accounts.
\newblock In \emph{Proceedings of the International AAAI Conference on Web and Social Media}, volume~17, 638--649.

\bibitem[{Minnich et~al.(2017)Minnich, Chavoshi, Koutra, and Mueen}]{minnich2017botwalk}
Minnich, A.; Chavoshi, N.; Koutra, D.; and Mueen, A. 2017.
\newblock BotWalk: Efficient adaptive exploration of Twitter bot networks.
\newblock In \emph{Proceedings of the 2017 IEEE/ACM international conference on advances in social networks analysis and mining 2017}, 467--474.

\bibitem[{Morris, Kersting, and Mutzel(2017)}]{morris2017glocalized}
Morris, C.; Kersting, K.; and Mutzel, P. 2017.
\newblock Glocalized weisfeiler-lehman graph kernels: Global-local feature maps of graphs.
\newblock In \emph{2017 IEEE International Conference on Data Mining (ICDM)}, 327--336. IEEE.

\bibitem[{Murtfeldt et~al.(2024)Murtfeldt, Alterman, Kahveci, and West}]{murtfeldt2024rip}
Murtfeldt, R.; Alterman, N.; Kahveci, I.; and West, J.~D. 2024.
\newblock RIP Twitter API: A eulogy to its vast research contributions.
\newblock \emph{arXiv preprint arXiv:2404.07340}.

\bibitem[{Shervashidze et~al.(2011)Shervashidze, Schweitzer, Van~Leeuwen, Mehlhorn, and Borgwardt}]{shervashidze2011weisfeiler}
Shervashidze, N.; Schweitzer, P.; Van~Leeuwen, E.~J.; Mehlhorn, K.; and Borgwardt, K.~M. 2011.
\newblock Weisfeiler-lehman graph kernels.
\newblock \emph{Journal of Machine Learning Research}, 12(9).

\bibitem[{Shervashidze et~al.(2009)Shervashidze, Vishwanathan, Petri, Mehlhorn, and Borgwardt}]{shervashidze2009efficient}
Shervashidze, N.; Vishwanathan, S.; Petri, T.; Mehlhorn, K.; and Borgwardt, K. 2009.
\newblock Efficient graphlet kernels for large graph comparison.
\newblock In \emph{Artificial intelligence and statistics}, 488--495. PMLR.

\bibitem[{Sugiyama and Borgwardt(2015)}]{sugiyama2015halting}
Sugiyama, M.; and Borgwardt, K. 2015.
\newblock Halting in random walk kernels.
\newblock \emph{Advances in neural information processing systems}, 28.

\bibitem[{Tardelli et~al.(2022)Tardelli, Avvenuti, Tesconi, and Cresci}]{tardelli2022detecting}
Tardelli, S.; Avvenuti, M.; Tesconi, M.; and Cresci, S. 2022.
\newblock Detecting inorganic financial campaigns on Twitter.
\newblock \emph{Information Systems}, 103: 101769.

\bibitem[{Topping et~al.(2021)Topping, Di~Giovanni, Chamberlain, Dong, and Bronstein}]{topping2021understanding}
Topping, J.; Di~Giovanni, F.; Chamberlain, B.~P.; Dong, X.; and Bronstein, M.~M. 2021.
\newblock Understanding over-squashing and bottlenecks on graphs via curvature.
\newblock \emph{arXiv preprint arXiv:2111.14522}.

\bibitem[{Tsitsulin et~al.(2018)Tsitsulin, Mottin, Karras, Bronstein, and M{\"u}ller}]{tsitsulin2018netlsd}
Tsitsulin, A.; Mottin, D.; Karras, P.; Bronstein, A.; and M{\"u}ller, E. 2018.
\newblock Netlsd: hearing the shape of a graph.
\newblock In \emph{Proceedings of the 24th ACM SIGKDD International Conference on Knowledge Discovery \& Data Mining}, 2347--2356.

\bibitem[{Varol et~al.(2017)Varol, Ferrara, Menczer, and Flammini}]{varol2017early}
Varol, O.; Ferrara, E.; Menczer, F.; and Flammini, A. 2017.
\newblock Early detection of promoted campaigns on social media.
\newblock \emph{EPJ data science}, 6: 1--19.

\bibitem[{Veli{\v{c}}kovi{\'c} et~al.(2018)Veli{\v{c}}kovi{\'c}, Cucurull, Casanova, Romero, Li{\`o}, and Bengio}]{velivckovic2018graph}
Veli{\v{c}}kovi{\'c}, P.; Cucurull, G.; Casanova, A.; Romero, A.; Li{\`o}, P.; and Bengio, Y. 2018.
\newblock Graph Attention Networks.
\newblock In \emph{International Conference on Learning Representations}.

\bibitem[{Wilkinson et~al.(2016)Wilkinson, Dumontier, Aalbersberg, Appleton, Axton, Baak, Blomberg, Boiten, da~Silva~Santos, Bourne et~al.}]{fair}
Wilkinson, M.~D.; Dumontier, M.; Aalbersberg, I.~J.; Appleton, G.; Axton, M.; Baak, A.; Blomberg, N.; Boiten, J.-W.; da~Silva~Santos, L.~B.; Bourne, P.~E.; et~al. 2016.
\newblock The FAIR Guiding Principles for scientific data management and stewardship.
\newblock \emph{Scientific data}, 3(1): 1--9.

\bibitem[{Wu et~al.(2023)Wu, Shi, Wang, Zeng, and Sun}]{wu23}
Wu, Y.; Shi, J.; Wang, P.; Zeng, D.; and Sun, C. 2023.
\newblock DeepCatra: Learning flow-and graph-based behaviours for Android malware detection.
\newblock \emph{IET Information Security}, 17(1): 118--130.

\bibitem[{Xu et~al.(2018)Xu, Hu, Leskovec, and Jegelka}]{xu2018powerful}
Xu, K.; Hu, W.; Leskovec, J.; and Jegelka, S. 2018.
\newblock How powerful are graph neural networks?
\newblock \emph{arXiv preprint arXiv:1810.00826}.

\bibitem[{Yanardag and Vishwanathan(2015)}]{IMDB}
Yanardag, P.; and Vishwanathan, S. 2015.
\newblock Deep Graph Kernels.
\newblock In \emph{Proceedings of the 21th ACM SIGKDD International Conference on Knowledge Discovery and Data Mining}, KDD '15, 1365–1374. New York, NY, USA: Association for Computing Machinery.
\newblock ISBN 9781450336642.

\bibitem[{Ying et~al.(2018)Ying, You, Morris, Ren, Hamilton, and Leskovec}]{ying2018hierarchical}
Ying, Z.; You, J.; Morris, C.; Ren, X.; Hamilton, W.; and Leskovec, J. 2018.
\newblock Hierarchical graph representation learning with differentiable pooling.
\newblock \emph{Advances in neural information processing systems}, 31.

\bibitem[{You et~al.(2020)You, Chen, Sui, Chen, Wang, and Shen}]{you2020graph}
You, Y.; Chen, T.; Sui, Y.; Chen, T.; Wang, Z.; and Shen, Y. 2020.
\newblock Graph contrastive learning with augmentations.
\newblock \emph{Advances in neural information processing systems}, 33: 5812--5823.

\bibitem[{Zannettou et~al.(2019)Zannettou, Caulfield, De~Cristofaro, Sirivianos, Stringhini, and Blackburn}]{zannettou2019disinformation}
Zannettou, S.; Caulfield, T.; De~Cristofaro, E.; Sirivianos, M.; Stringhini, G.; and Blackburn, J. 2019.
\newblock Disinformation warfare: Understanding state-sponsored trolls on Twitter and their influence on the web.
\newblock In \emph{Companion proceedings of the 2019 world wide web conference}, 218--226.

\bibitem[{Zhang et~al.(2018)Zhang, Cui, Neumann, and Chen}]{zhang2018end}
Zhang, M.; Cui, Z.; Neumann, M.; and Chen, Y. 2018.
\newblock An end-to-end deep learning architecture for graph classification.
\newblock In \emph{Proceedings of the AAAI conference on artificial intelligence}.

\end{thebibliography}

\section{Ethics Checklist}

\begin{enumerate}

\item For most authors...
\begin{enumerate}
    \item  Would answering this research question advance science without violating social contracts, such as violating privacy norms, perpetuating unfair profiling, exacerbating the socio-economic divide, or implying disrespect to societies or cultures?
    \answerYes{Yes}
  \item Do your main claims in the abstract and introduction accurately reflect the paper's contributions and scope?
    \answerYes{Yes}
   \item Do you clarify how the proposed methodological approach is appropriate for the claims made? 
    \answerYes{Yes. This is mentioned in section titled Engagement networks: campaign or not.}
   \item Do you clarify what are possible artifacts in the data used, given population-specific distributions?
    \answerYes{Yes. This is mentioned in Campaigns collection methodology, Non-campaign collection methodology, and Building networks.}
  \item Did you describe the limitations of your work?
    \answerYes{Yes. In section titled Limitations.}
  \item Did you discuss any potential negative societal impacts of your work?
    \answerNo{No. We do not foresee our contributions to have any negative societal impacts on their own.}
      \item Did you discuss any potential misuse of your work?
    \answerNo{No. To the best of our knowledge, there are no known instances of misuse related to our work.}
    \item Did you describe steps taken to prevent or mitigate potential negative outcomes of the research, such as data and model documentation, data anonymization, responsible release, access control, and the reproducibility of findings?
    \answerYes{Yes. We do anonimize the user names while creating the graphs. This is done on the users, their tweets and their captions.}
  \item Have you read the ethics review guidelines and ensured that your paper conforms to them?
    \answerYes{Yes}
\end{enumerate}

\item Additionally, if your study involves hypotheses testing...
\begin{enumerate}
  \item Did you clearly state the assumptions underlying all theoretical results?
    \answerNo{No. We don't have any theoretical reasons and therefore haven't stated any assumptions}
  \item Have you provided justifications for all theoretical results?
    \answerNo{No. We don't have any theoretical results.}
  \item Did you discuss competing hypotheses or theories that might challenge or complement your theoretical results?
    \answerNo{No. We don't have any competing hypothesis'.}
  \item Have you considered alternative mechanisms or explanations that might account for the same outcomes observed in your study?
    \answerNo{No. We don't do any hypothesis testing.}
  \item Did you address potential biases or limitations in your theoretical framework?
    \answerNo{No. We don't have any theoretical limitations}
  \item Have you related your theoretical results to the existing literature in social science?
    \answerNo{No. We don't have any theoretical results in the paper.}
  \item Did you discuss the implications of your theoretical results for policy, practice, or further research in the social science domain?
    \answerNo{No. We don't have any theoretical results in the paper.}
\end{enumerate}

\item Additionally, if you are including theoretical proofs...
\begin{enumerate}
  \item Did you state the full set of assumptions of all theoretical results?
    \answerNo{No. We don't have any theoretical results in the paper.}
	\item Did you include complete proofs of all theoretical results?
    \answerNo{No. We don't have any theoretical results in the paper.}
\end{enumerate}

\item Additionally, if you ran machine learning experiments...
\begin{enumerate}
  \item Did you include the code, data, and instructions needed to reproduce the main experimental results (either in the supplemental material or as a URL)?
    \answerYes{Yes. The URL for the code and the accompanying instructions are given in Graph classification on engagement networks. The data is provided in the URL in the abstract.}
  \item Did you specify all the training details (e.g., data splits, hyperparameters, how they were chosen)?
    \answerYes{Yes. All of them are specified in the section titled Graph classification on engagement networks under the subsection titled Experimental setup and Graph classifiers.}
     \item Did you report error bars (e.g., with respect to the random seed after running experiments multiple times)?
    \answerYes{Yes. This is observable in Table 4, 5 and 6.}
	\item Did you include the total amount of compute and the type of resources used (e.g., type of GPUs, internal cluster, or cloud provider)?
    \answerYes{Yes. This is specified in the section titled Graph classification on engagement networks under the subsection titled Experimental setup.}
     \item Do you justify how the proposed evaluation is sufficient and appropriate to the claims made? 
    \answerYes{Yes. We do specify that in the section titled Graph classification on engagement networks.}
     \item Do you discuss what is ``the cost`` of misclassification and fault (in)tolerance?
    \answerNo{We do not. The main objective of the paper is providing a challenging dataset, not a new method.}
  
\end{enumerate}

\item Additionally, if you are using existing assets (e.g., code, data, models) or curating/releasing new assets, \textbf{without compromising anonymity}...
\begin{enumerate}
  \item If your work uses existing assets, did you cite the creators?
    \answerYes{Yes we do}
  \item Did you mention the license of the assets?
    \answerYes{Yes. We have mentioned the license of our datasets?}
  \item Did you include any new assets in the supplemental material or as a URL?
    \answerNo{No. We dont have any new assets.}
  \item Did you discuss whether and how consent was obtained from people whose data you're using/curating?
    \answerYes{Yes. The data was obtained using the Twitter API before it became a payed feature.}
  \item Did you discuss whether the data you are using/curating contains personally identifiable information or offensive content?
    \answerYes{Yes. We did discus this in the ethics section.}
    \item If you are curating or releasing new datasets, did you discuss how you intend to make your datasets FAIR (see \citep{fair})?
    \answerYes{We do provide a rich amount of metadata and our data is accessible. We also made an effort to keep our data interoperable and re-usable.}
    \item If you are curating or releasing new datasets, did you create a Datasheet for the Dataset (see \citep{gebru2021datasheets})? 
    \answerYes{We have the datasheet included in the appendix.}
\end{enumerate}

\item Additionally, if you used crowdsourcing or conducted research with human subjects, \textbf{without compromising anonymity}...
\begin{enumerate}
  \item Did you include the full text of instructions given to participants and screenshots? 
  \answerNA{NA}
  \item Did you describe any potential participant risks, with mentions of Institutional Review Board (IRB) approvals?
    \answerNA{NA}
  \item Did you include the estimated hourly wage paid to participants and the total amount spent on participant compensation?
    \answerNA{NA.} 
   \item Did you discuss how data is stored, shared, and de-identified?
    \answerNA{NA}
\end{enumerate}

\end{enumerate}

\section{Author statement}
The dataset is released under a CC-BY license, enabling free sharing and adaptation for research or development purpose. We bear all responsibility in case of violation of rights.

\appendix

\section{Appendix}

\begin{figure*}[htbp]
    \centering
    \includegraphics[width=.9\textwidth]{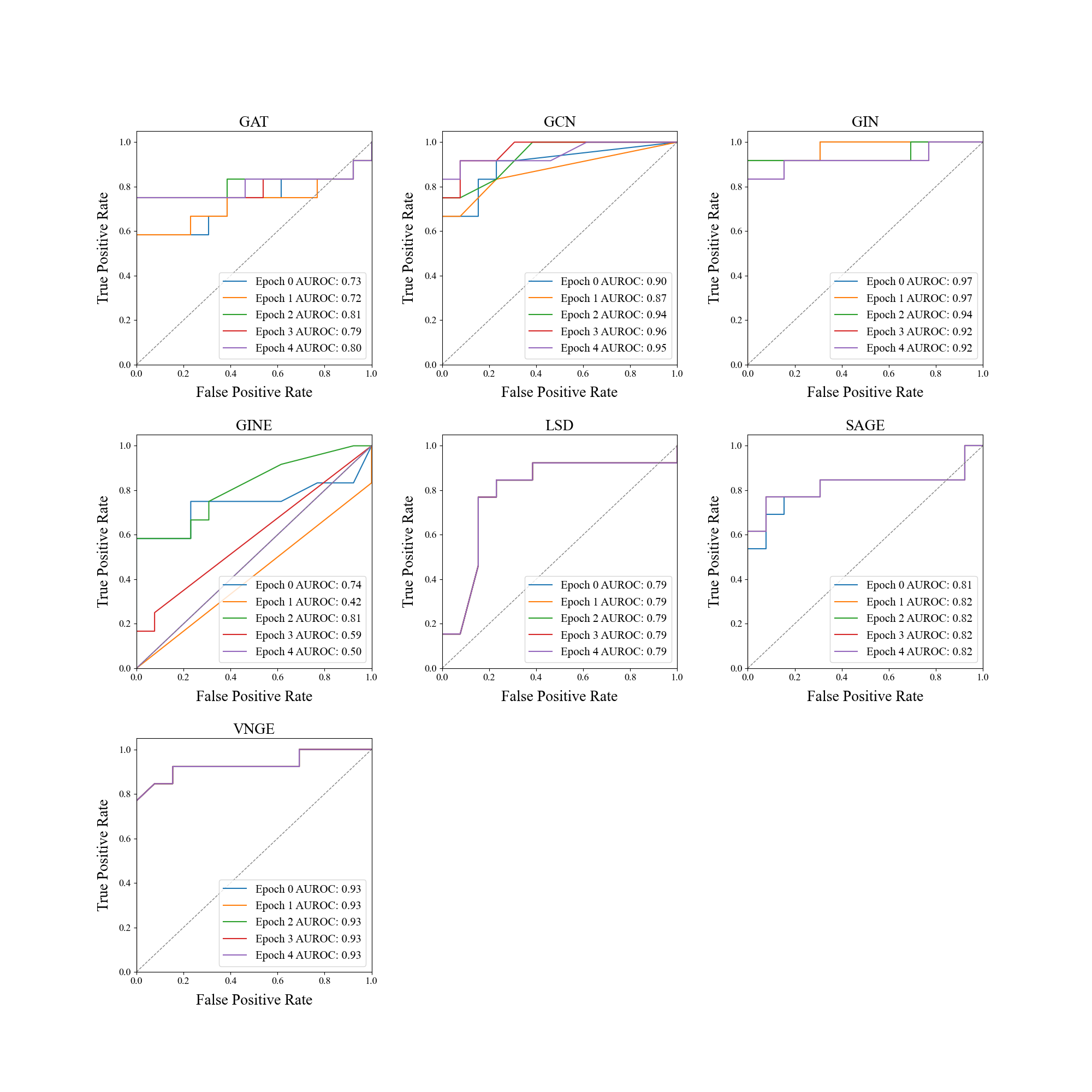} 
    \caption{Receiver Operating Characteristic (ROC) curves for campaign vs. non-campaign classification across the small dataset.}
    \label{fig:ROC_small}
\end{figure*}

\begin{figure*}[htbp]
    \centering
    \includegraphics[width=\textwidth]{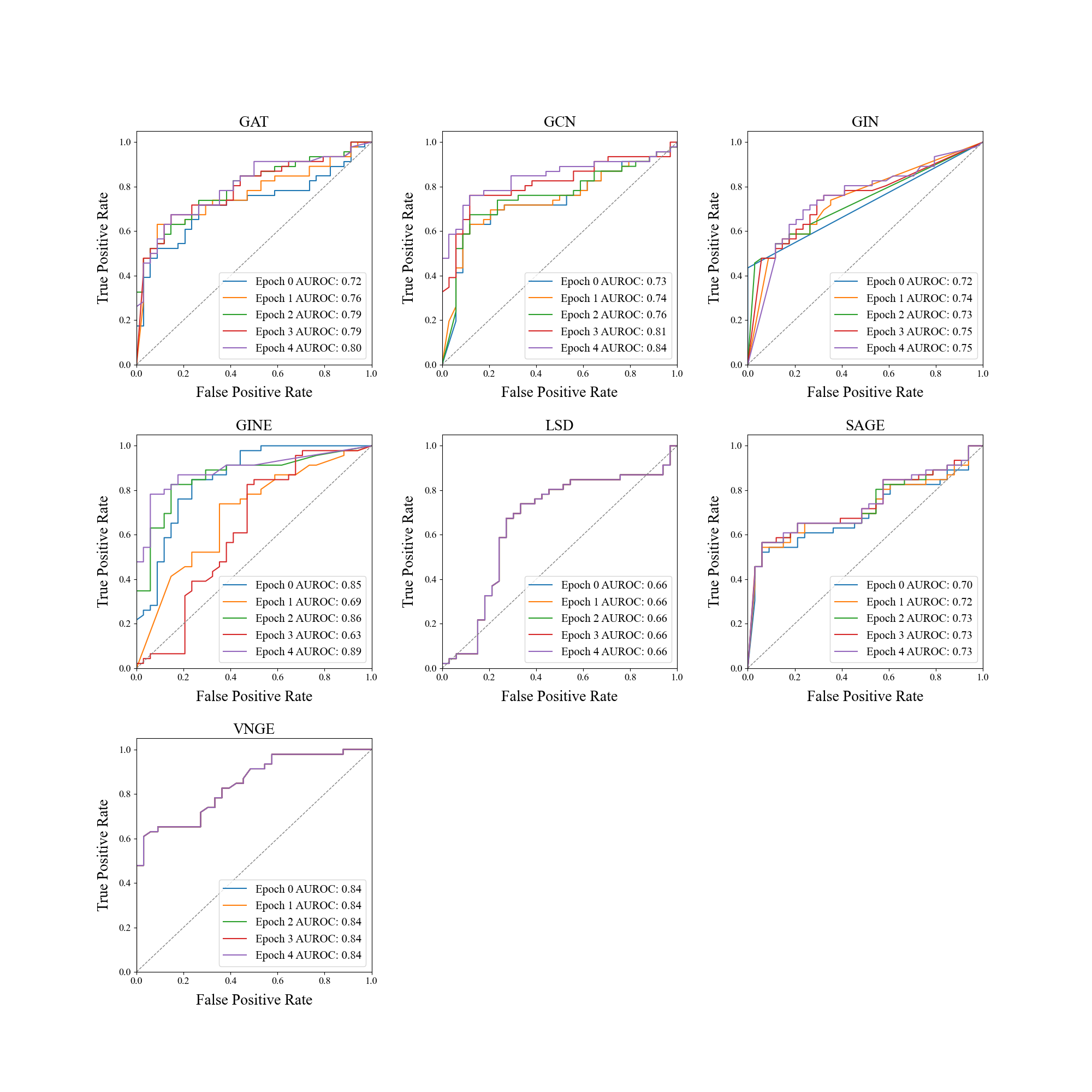} 
    \caption{Receiver Operating Characteristic (ROC) curves for campaign vs. non-campaign classification across the complete dataset.}
    \label{fig:ROC_all}
\end{figure*}

\begin{table}[htbp]
    \setlength{\tabcolsep}{1mm} 
    \fontsize{9pt}{11pt}\selectfont
    \centering
    \begin{tabular}{l|l|r|r|r|r|r|r}
    & \fontsize{10}{12}\selectfont \textbf{Sub-type} & \multicolumn{3}{|c|}{\fontsize{10}{12}\selectfont \textbf{\# of Conn. Comp.}} & \multicolumn{3}{|c}{\fontsize{10}{12}\selectfont \textbf{fLCC}}\\
    
    \hline
    & & \fontsize{10}{12}\selectfont \textbf{Min} & \fontsize{10}{12}\selectfont \textbf{Max} & \fontsize{10}{12}\selectfont \textbf{Avg} & \fontsize{10}{12}\selectfont \textbf{Min} & \fontsize{10}{12}\selectfont \textbf{Max} & \fontsize{10}{12}\selectfont \textbf{Avg}\\
 
    \hline
     \multirow{6}{*}{\rotatebox[origin=c]{90}{Campaign}} &       Politics        &                                                      1 & 2,004 & 207.29 & 0.355                                                      & 1 & 0.800\\
                                                        &        Reform          & 1 & 112 & 13.16 & 0.396 & 1 & 0.826\\
                                                        &        News            & 17 & 2,138 & 578.67 & 0.147 & 0.975 & 0.735\\
                                                        &        Finance         & 6 & 1,486 & 159.71 & 0.257 & 0.973 & 0.691\\
                                                        &        Noise           & 16 & 8,908 & 1,865.22 & 0.065 & 0.976 & 0.469\\
                                                        &        Cult            & 12 & 122 & 67.00 & 0.293 & 0.899 & 0.553\\   
                                                        & \textbf{Overall} & \textbf{1} & \textbf{8,908} & \textbf{269.47} & \textbf{0.065} & \textbf{1} & \textbf{0.767}\\
    
    \hline
    \multirow{7}{*}{\rotatebox[origin=c]{90}{Non-Campaign}} &                News            & 10 & 818 & 6,169 & 0.203 & 0.989 & 0.793\\
    &                Sports          & 54 & 3,114 & 576.00 & 0.180 & 0.981 & 0.655\\
    &                Festival        & 128 & 7,289 & 1,721.24 & 0.349 & 0.924 &                                     0.793\\
    &                Internal        & 164 & 7,605 & 1,096.45 & 0.337 & 0.988 &                                     0.793 \\
    &                Common          & 103 & 1,851 & 788.13 & 0.298 & 0.940 &                                       0.945\\
    &                Enter.   & 101 & 396 & 193.28 & 0.570 & 0.953 & 0.792\\
    &                Sp. cam.     & 68 & 105 & 76.00 & 0.885 & 0.926 & 0.906\\
    & \textbf{Overall} & \textbf{54} & \textbf{7,605} & \textbf{675.80} & \textbf{0.180} & \textbf{0.989} & \textbf{0.816}\\
    \end{tabular}
    \caption{Description of connected components in the graph. Here fLCC is the fraction of the largest connected component to the whole graph.}
    \label{tab:connected}
\end{table}


\newpage

\end{document}